\documentclass[fleqn,12pt]{wlscirep}
\usepackage[utf8]{inputenc}
\usepackage[T1]{fontenc}
\usepackage{chemformula}
\let\ce\ch
\usepackage{adjustbox}
\usepackage{lipsum}
\usepackage{mathtools}
\usepackage{physics}
\usepackage{floatrow}
\usepackage{hyperref}
\usepackage[justification=centering]{caption}
\usepackage{amsmath}
\usepackage{nccmath}
\usepackage{subcaption}
\usepackage{floatrow}
\usepackage[thinlines]{easytable}
\usepackage{cleveref}
\usepackage{gensymb}
\usepackage{booktabs}  
\usepackage{ltablex}
 \usepackage{bbm}

\title{Magnetic field effects in biology from the perspective of the radical pair mechanism}

\author[1,2,3,*]{Hadi Zadeh-Haghighi}
\author[1,2,3,*]{Christoph Simon}

\affil[1]{Department of Physics and Astronomy, University of Calgary, Calgary, AB, T2N 1N4, Canada}
\affil[2]{Institute for Quantum Science and Technology, University of Calgary, Calgary, AB, T2N 1N4, Canada}
\affil[3]{Hotchkiss Brain Institute, University of Calgary, Calgary, AB, T2N 1N4, Canada}

\affil[
*]
{hadi.zadehhaghighi@ucalgary.ca \& csimo@ucalgary.ca}

\begin{abstract}
A large and growing body of research shows that weak magnetic fields can significantly influence various biological systems, including plants, animals, and humans. However, the underlying mechanisms behind these phenomena remain elusive. It is remarkable that the magnetic energies implicated in these effects are much smaller than thermal energies. Here we review these observations, of which there are now hundreds, and we suggest that a viable explanation is provided by the radical pair mechanism, which involves the quantum dynamics of the electron and nuclear spins of naturally occurring transient radical molecules. While the radical pair mechanism has been studied in detail in the context of avian magnetoreception, the studies reviewed here show that magnetosensitivity is widespread throughout biology. We review magnetic field effects on various physiological functions, organizing them based on the type of the applied magnetic fields, namely static, hypomagnetic, and oscillating magnetic fields, as well as isotope effects. We then review the radical pair mechanism as a potential unifying model for the described magnetic field effects, and we discuss plausible candidate molecules that might constitute the radical pairs. We review recent studies proposing that the quantum nature of the radical pairs provides promising explanations for xenon anesthesia, lithium effects on hyperactivity, magnetic field and lithium effects on the circadian clock, and hypomagnetic field effects on neurogenesis and microtubule assembly. We conclude by discussing future lines of investigation in this exciting new area of quantum biology related to weak magnetic field effects. \\


\end{abstract}

\begin{document}

\maketitle

\tableofcontents

\section{\label{sec:level1} Introduction}

Sensitivity to weak magnetic fields is abundant throughout biology, as discussed in numerous review articles \cite{KETCHEN1978,2018,Jones2016,Dini2005,Albuquerque2016,Fan2021,shupak2003therapeutic,Marycz2018,Gartzke2002,McKay2007,Markov2007,Davanipour2009,Radhakrishnan2019,Saunders2005,Wang2017ros,Vergallo2018,Nyakane2019,Sarraf2020,Maffei2014,Villa1991,marino1977biological,Binhi2017,Zhang2017book,Xue2021,Mo2014Transcriptome}. Effects of either static or oscillating weak magnetic fields have been reported on the circadian clock, electron transfer in cryptochrome, stem cells, calcium concentration, the brain's functions such as action potentials, reactive oxygen species, development, neuronal activities, DNA, memory, anxiety, analgaesia, genetics, and many other functions (See Section \ref{sec:MFE}). Despite the wealth of observations, thus far there is no clear explanation for the mechanism behind these phenomena. This is mainly due to the fact that the corresponding energies for such effects are far smaller than thermal energies. 

However, there is a promising quantum physics (or spin chemistry) concept that can account for the effects of such weak fields, namely the radical pair mechanism \cite{Schulten1978,Steiner1989}. This mechanism, which is an example of the emerging field of quantum biology \cite{Mohseni2009,Lambert2012,Ball2011,Cao2020,Kim2021}, has been studied in significant detail in the comparatively narrow context of bird magnetoreception \cite{Wiltschko1995,Wiltschko1972,Cochran2004,Zapka2009,Wiltschko2010,Mouritsen2018,Wiltschko2021,Mouritsen2022}, where it is accepted as one of the leading potential explanations for how birds sense magnetic fields, and in particular the Earth's magnetic field, for the purpose of navigation. 

\par


 It involves magnetically sensitive intermediate molecules, so-called radical pairs \cite{Grissom1995,Rodgers2009,Hiscock2016,Hore2016,Xu2021,Wong2021,Schulten1978}. The key ingredient is the spin correlation between two unpaired electrons, one on the donor molecule and the other on the acceptor molecule. Depending on the initial spin configuration of the donor molecule, this initial spin correlation of the radical pair will be either a singlet (S) or a triplet (T) state, which are respectively spin-$0$ and spin-$1$ states (See Section \ref{sec:spin-RP} for further discussion). Due to the spin interactions with its environment (in particular with external magnetic fields and with nearby nuclear spins), the state of the radical pair will oscillate between S and T states \cite{Steiner1989,Timmel1998}. Each spin state, S and T, can lead to different reaction products, providing an example of spin chemistry \cite{Schulten1976,Hore2020}. The energies induced by the above-mentioned magnetic fields are hundreds of thousands of times smaller than thermal energies, $k_B T$ ($k_B$ is Boltzmann constant and $T$ is temperature), which are associated with motions, rotation, and vibrations in biological environments. In thermal equilibrium, the energies required to alter the rate or yield of a chemical transformation should be at least comparable to $K_B T$. Due to this, the radical pair mechanism was originally ignored in the context of physiology. However, the situation differs in systems far from thermal equilibrium, which is the case for radical pairs \cite{Hore2016}. Sensitivity to weak magnetic fields is one of the key properties of radical pair reactions. Nowadays, many research labs study the role of radical pairs in (bio)chemical reactions \cite{Steiner1989,Woodward2002,Rodgers2009m,Hore2020,Zhang2020cata,Beretta2019}. \par

Recent studies have proposed roles for radical pairs beyond avian magnetoreception, in particular in xenon-induced anesthesia \cite{Smith2021}, lithium effects on mania \cite{Zadeh2021Li}, magnetic field and lithium effects on the circadian clock \cite{Zadeh2022CC}, and hypomagnetic field effects on microtubule reorganization \cite{ZadehHaghighi2022} and neurogenesis \cite{rishabh2021radical} (where hypomagnetic fields are fields much weaker than that of the Earth). Here we suggest that the radical pair mechanism is in fact quite common in biology, and that it may provide an explanation for many of the weak magnetic field effects on physiological functions that have been observed.\par 

This paper, which is part review and part perspective article, is organised as follows. Section \ref{sec:MFE} briefly surveys studies reporting effects of low-intensity magnetic fields on biological systems, including effects of static (Section \ref{sec:SMF}), hypomagnetic (Section \ref{sec:HMFE}), and oscillating (Section \ref{sec:OMF}) magnetic fields. We further survey studies on isotope effects in biology form a spin perspective. In Section \ref{sec:RPM}, we discuss how the radical pair mechanism can account for static, hypomagnetic, and oscillating magnetic field effects. Section \ref{sec:RP-Candidates} reviews possible candidate molecules for radical pair formation in biological systems. In Section \ref{sec:RPM-Brain}, we review the above-mentioned recent studies on the possible biological roles of radical pairs beyond avian magnetoreception. Section \ref{sec:Remarks} discusses important directions for further investigation.

\section{Magnetosensitivity in biology} \label{sec:MFE}
There is a considerable amount of research investigating magnetic field effects on biological functions \cite{Repacholi1999,Galland2005,Pazur2007,Buchachenko2014,Zhang2017book,Lai2019,Guerra2019,Binhi2022,Bertea2015,Maffei2022}. In the following, we review the effects of low-intensity magnetic fields on biology. We organize this section based on the type of magnetic fields, namely static magnetic fields, hypomagnetic fields, and oscillating magnetic fields. Isotope effects in biology, which can be related to nuclear magnetic moments, are also discussed at the end of this section.

\subsection{Static magnetic field} \label{sec:SMF}

\subsubsection{Cryptochrome}

In the context of avian magnetoreception in animals, the canonical proteins are cryptochromes \cite{Hore2016,Xu2021}. Maeda et al. demonstrated that photo-induced flavin—tryptophan radical pairs in cryptochrome is magnetically sensitive \cite{Maeda2012}.  Moreover, Ahmad et al. observed that hypocotyl growth inhibition in higher plants are sensitive to the magnetic field, where such responses are linked to cryptochrome-dependent signaling pathways \cite{Ahmad2006}. Sheppard et al. reported that magnetic fields of a few mT could influence photo-induced electron transfer reactions in \textit{Drosophila} cryptochrome \cite{Sheppard2017}. Further, Mayer et al. showed that a static magnetic field of 100 mT substantially affected seizure response in \textit{Drosophila} larvae in a cryptochrome-dependent manner \cite{Marley2014}. In addition, using a transgenic approach, Foley et al. showed that human cryptochrome-2 has the molecular capability to function as a light-sensitive magnetosensor \cite{Foley2011}. Applying a 0.5 mT magnetic field, Ahmad and co-workers reported that cryptochrome responses were enhanced by the magnetic field, including dark-state processes following the cryptochrome photoreduction step \cite{Pooam2018,Hammad2020}. Further, there have been extensive studies on the radical pair mechanism for cryptochrome(s) \cite{Hiscock2016,Hore2016}. Table \ref{tab:SMF} summarizes static magnetic field effects on various biological functions.

\subsubsection{Genetics}

It is known that exposure to magnetic fields has genetic consequences \cite{McCann1998}. Giorgi et al. showed that chronic exposure to magnetic fields (0.4-0.7 mT) increased the body size and induced lethal mutations in populations of \textit{Drosophila melanogaster} \cite{Giorgi1992}. Furthermore, a magnetic field of 35 mT decreased the wing size in \textit{Drosophila melanogaster} \cite{StamenkoviRadak2001} (See Table \ref{tab:SMF}).

\subsubsection{Circadian clock} \label{sec:SMF-CC}
It has been shown that magnetic fields can modulate the circadian clock \cite{Close2014,Close2014b,Bradlaugh2021}. Yoshii et al. \cite{yoshii2009cryptochrome} showed that the effects of static magnetic fields affected the circadian clock of \textit{Drosophila} and reported that exposure to these fields slowed down the clock rhythms in the presence of blue light, with a maximal change at 300 $\mu$T, and reduced effects at both lower and slightly higher field strengths. We discuss this observation further from the perspective of the radical pair mechanism in Section \ref{sec:RPM-CC} (See Table \ref{tab:SMF}).

 \subsubsection{Stem cells}
Static magnetic field has been commonly used in medicine as a tool to increase wound healing, bone regeneration and as a component of magnetic resonance techniques. However, recent data shed light on deeper mechanism of Static magnetic field action on physiological properties of different cell populations, including stem cells. It is known that static magnetic fields can increase wound healing and bone regeneration \cite{Marycz2018}. Huizen et al. reported that weak magnetic fields ($<$1 mT) alter stem cell-mediated growth, where changes in ROS were implicated \cite{VanHuizen2019}. The authors suggested that the radical pair mechanism may be the potential explanation for their observations. Zheng et al. showed that static magnetic field of 1, 2, or 4 mT regulated proliferation, migration, and differentiation of human dental pulp stem cells \cite{Zheng2018}. It is also known that applied static magnetic fields (0.5–30 mT) affect stem cells \textit{in vitro} \cite{Tavasoli2009,JavaniJouni2013,Jouni2014} (See Table \ref{tab:SMF}).


\subsubsection{Calcium}
Fanelli et al. reported that magnetic fields allow the indefinite survival and replication of the cells hit by apoptogenic agents. The anti-apoptosis effect was found to be mediated by the ability of the fields to increase \ch{Ca^{2+}} influx from the extracellular medium. In that experiment, the geomagnetic field was not shielded. They found 0.6 mT to be the minimal intensity required to detect an anti-apoptotic effect \cite{FANELLI1999}. Moreover, it has been shown that weak static magnetic fields can influence myosin phosphorylation in a cell-free preparation in a \ch{Ca^{2+}}-dependent manner \cite{Markov1997}. Tenuzzo and colleagues observed that exposure to a 6 mT static magnetic field influenced \ch{Ca^{2+}} concentration and bcl-2, bax, p53 and hsp70 expression in freshly isolated and \textit{in vitro} aged human lymphocytes \cite{Tenuzzo2009}. Further, Chionna et al. showed that exposure to a static magnetic field of 6 mT of Hep G2 cells resulted in time dependent modifications in cell shape, cell surface, sugar residues, cytoskeleton, and apoptosis \cite{Chionna2005}. They reported that after 24 h exposure, the cells had a less flat shape due to partial detachment from the culture dishes. They further observed that microfilaments and microtubules were modified in a time dependent manner. They also suggested that the induced apoptosis was likely due to the increment of \ch{Ca^{2+}} during exposure. In another study, Tenuzzo and co-workers showed that cell viability, proliferation, intracellular \ch{Ca^{2+}} concentration and morphology in several primary cultures and cell lines can be influenced by a 6 mT magnetic field \cite{Tenuzzo2006} (See Table \ref{tab:SMF}).

\subsubsection{Neurons and brain}
Exposure to static magnetic fields can have impacts on various brain functions. McLean et al. reported that a static magnetic field in the 10 mT range blocked sensory neuron action potentials in the somata of adult mouse dorsal root ganglion neurons in monolayer dissociated cell culture \cite{McLean1995}. It has also been shown that exposure to a transcranial static magnetic field over the supplementary motor area can modulate resting-state activity and motor behavior associated with modulation of both local and distant functionally-connected cortical circuits \cite{PinedaPardo2019}. Static magnetic field exposure can also affect the production of melatonin \cite{Welker1983,Reiter1992,Reiter1993,Reiter1995}, the pineal gland \cite{Semm1980,Lerchl1991}, and cause functional alterations in immature cultured rat hippocampal neurons \cite{Hirai2003}. Further, Dileone et al. observed that an applied transcranial static magnetic field can induce dopamine-dependent changes of cortical excitability in patients with Parkinson’s disease \cite{Dileone2017}. In addition, neuron firing frequency can also be affected by static magnetic field intensity \cite{Azanza1995,Spasi2011}. There exist a considerable number of studies indicating the effects of applied magnetic field on pain sensitivity (nociception) and pain inhibition (analgesia) \cite{DelSeppia2007}. Additionally, it has been known that a static magnetic field (50 mT) can influence symptomatic diabetic neuropathy \cite{Weintraub2003} (See Table \ref{tab:SMF}).

\subsubsection{Reactive oxygen species} \label{sec:SMF-ROS}

Reactive oxygen species (ROS) is the collection of derivatives of molecular oxygen that occur in biology, which can be categorized into two types, free radicals and non-radical species. The non-radical species are hydrogen peroxide (\ch{H2O2}), organic hydroperoxides (\ch{ROOH}), singlet molecular oxygen (\ch{^1O2}), electronically excited carbonyl, ozone (\ch{O3}), hypochlorous acid (\ch{HOCl}, and hypobromous acid \ch{HOBr}). Free radical species are superoxide anion radical (\ch{O2^{.-}}), hydroxyl radical (\ch{·OH}), peroxyl radical (\ch{ROO·}), and alkoxyl radical (\ch{RO·}) \cite{Sies2020}. Any imbalance of ROS can lead to adverse effects. \ch{H2O2} and \ch{O2^{.-}} are the main redox signaling agents. It is now well known that ROS are essential for physiology as functional signalling entities. \ch{H2O2} plays a crucial role in redox regulation of biological functions, where its intracellular concentration is under tight control. The cellular concentration of \ch{H2O2} is about 10$^{-8}$ M, which is almost a thousand times more than that off \ch{O2^{.-}}. Transmembrane NADPH oxidases (NOXs) \cite{Bedard2007,Moghadam2021} and the mitochondrial electron transport chain (ETC) \cite{Murphy2008,HernansanzAgustn2021} are the major sources of \ch{O2^{.-}} and \ch{H2O2}. \par 

In a considerable number of studies, magnetic field effects in biology are accompanied with oxidative stress \cite{Okano2008,Ghodbane2013,Wang2017ros}, which is \textit{``an imbalance between oxidants and antioxidants in favor of the oxidants, leading to a disruption of redox signaling and control and/or molecular damage."} \cite{Sies2017,Sies2017a,Sies2020a}. Studies found that exposure to static magnetic fields of 2.2 mT \cite{Calabr2013} and  31.7–232 mT \cite{Vergallo2014} increased the intercellular ROS in human neuroblastoma cells. Furthermore, De Nicola et al. observed that the intracellular ROS level in human monocyte tumor cells was raised when exposed to a static magnetic field \cite{DENICOLA2006}. Further, Bekhite et al. reported that static magnetic field exposure (1-10 mT) increased the \ch{H2O2} level in embryoid bodies \cite{Bekhite2010}. Later, the same group found an induced increase of ROS in cardiac progenitor cells derived from mouse cells by 0.2–5 mT static magnetic field, where ROS was suggested to be generated by NADPH oxidase \cite{Bekhite2013}. Sullivan et al. reported that 230–250 mT of a magnetic field elevated \ch{H2O2} in diploid embryonic lung fibroblast cell \cite{Sullivan2010}. Upon exposure to 45-60 $\mu$T, Marino and Castello observed an increase of \ch{H2O2} in the human fibrosarcoma cancer cell, which can be suppressed by reducing the geomagnetic field's strength \cite{Martino2011}. Further studies show that exposure to 60 mT magnetic field increased \ch{H2O2} production of human peripheral blood neutrophils \cite{Poniedziaek2013}. It has also been reported that the effects of an applied magnetic field of 10 mT on DOXO-induced toxicity and proliferation rate of cancer cells are correlated to ROS levels \cite{HajipourVerdom2018}. Furthermore, Carter et al. observed that a 3 mT static magnetic field can influence type 2 diabetes via regulating cellular ROS \cite{Carter2020,Yu2021}. Pooam et al. showed that applying low intensity static magnetic field modulated ROS generation in HEK293 cells. The authors suggested that the radical pair mechanism may explain that observation \cite{Pooam2020}. In a recent work, Sheu and co-workers reported that static low intensity magnetic field can regulate mitochondrial electron transport chain activity as thus enhance mitochondrial respiration.\cite{Sheu2022}. They observed that exposure to magnetic fields of 0-1.93 mT of mitochondria isolated from adult rat hearts produced a bell-shape increase in the respiratory control ratio with a maximum at 0.50 mT and a return to baseline at 1.50 mT. It was further observed that the magnetic field affected only the activity of the complexes 2, 3 and 5 but not 1 of the mitochondrial electron transport chain and several enzymes of the tricarboxylic acid cycle. The authors suggested that the low intensity magnetic field effects on the mitochondrial respiratory activity may be explained by the radical pair mechanism. Huizen and co-workers showed that weak magnetic fields ($<$1 mT) changed stem cell-mediated growth, where changes in ROS were implicated \cite{VanHuizen2019}.

\subsubsection{Others}

Ikeya et al. reported that exposure to magnetic field influenced autofluorescence in cells involving flavins \cite{Ikeya2021}. Studies also showed that static magnetic fields can affect the photoactivation reaction of E.coli DNA photolyase \cite{Henbest2008}. Moreover, Giachello et al. observed that applying static magnetic fields on blue-light activated cryptochromes in \textit{Drosophila} neurons resulted in an elevation of action potential firing \cite{Giachello2016}. Further, it is also known that the chemiluminescence intensity in Madin-Darby canine kidney cells is magnetic field dependent \cite{cheun2007biophoton}, where ROS are implicated. In solutions, flavin adenine dinucleotide is the key cofactor of cryptochrome. Antill and co-workers showed that flavin adenine dinucleotide photochemistry in solution is magnetic field sensitive ($<$20 mT) even at physiological pH and higher \cite{Antill2018}. 

Buchachenko et al. reported that applying 80 mT static magnetic field affected enzymatic ATP production \cite{Buchachenko2008}. Recently, Ercan et al. showed that exposure to magnetic fields (20, 42, 125, and 250 mT) can affect the magnetic properties, germination, chlorophyll fluorescence, and nutrient content of barley (\textit{Hordeum vulgare} L.) \cite{Ercan2022}. Further, it is observed that exposure to magnetic fields (10 and 30 mT) can deteriorate the antioxidant defense system of plant cells \cite{Sahebjamei2006}. Hao et al. reported that exposure to a 8.8 mT static magnetic field can enhance the killing effect of adriamycin on K562 cells \cite{Hao2010}. It is also observed that exposure to magnetic fields (2.9–4.6 mT) of soybean tissue culture enhances the regeneration and plant growth of shoot tips  \cite{Atak2007}. Teodori et al. showed that exposure of HL-60 cells to a 6 mT static magnetic field accelerated loss of integrity of plasma membrane during apoptosis \cite{Teodori2002}. It has been shown that exposure of human pro-monocytic U937 cells to a static magnetic field (6 mT) decreased the degree of macrophagic differentiation \cite{Pagliara2009}. Buemi et al. report that exposure to a 0.5 mT magnetic field of renal cell cultures and cortical astrocyte cultures from rats influenced cell proliferation and cell death balance \cite{Buemi2001}. They concluded that such magnetic field effects were cell type-dependent. It has been shown that exposure to magnetic fields (1 mT) significantly affected growth and sporulation of phytopathogenic microscopic fungi \cite{Nagy2004}.

Surma et al. found that the application of weak static magnetic field with intensities only a few times that of the geomagnetic field can accelerate the development of skeletal muscle cells, resulting in the formation of multinuclear hypertrophied myotubes \cite{Surma2014}. They further reported that these effects were accompanied by a 1.5- to 3.5-fold rise in the concentration of intracellular [\ch{Ca^{2+}}]$_i$.

\begin{table}[ht!]
\centering
\begin{adjustbox}{width=1\textwidth}
\begin{tabular}{lcr}
  \hline
 \textbf{System} & \textbf{Magnetic field}  & \textbf{References} \\ 
  \hline
 \textbf{Cryptochrome} &   &  \\   
cryptochrome responses enhanced & 0.5 mT & Pooam et al. (2018) \cite{Pooam2018}\\
cryptochrome responses enhanced & 0.5 mT & Hammad et al. (2020) \cite{Hammad2020}\\
seizure response in \textit{Drosophila} (cryptochrome-dependent) &Further,  100 mT  &Mayer et al. (2014)\cite{Marley2014}\\
photo-induced electron transfer reactions in \textit{Drosophila} cryptochrome& a few mT &Sheppard et al. (2017) \cite{Sheppard2017}\\
 
  \hline

body size increase and in  \textit{Drosophila melanogaster} &0.4-0.7 mT& Giorgi et al. (1992)\cite{Giorgi1992}\\

decrease in wing size in \textit{Drosophila melanogaster}&35 mT & Stamenkovi-Radak et al. (2001)\cite{StamenkoviRadak2001}\\

  \hline
\textbf{ Circadian clock} &   &  \\ 
circadian clock in \textit{Drosophila melanogaster} &$<$0.5 mT& Yoshii et al. (2009) \cite{yoshii2009cryptochrome}\\

  \hline
\textbf{Stem cell} &   &  \\ 
  
stem cell-mediated growth&$<$1 mT&Huizen et al. (2019) \cite{VanHuizen2019}\\
proliferation/migration/differentiation in human dental pulp stem cells&1/2/4 mT& Zheng et al. (2018)\cite{Zheng2018}\\
bone stem cells \textit{in vitro} & 0.5–30 mT & Abdolmaleki et al. \cite{Tavasoli2009,JavaniJouni2013,Jouni2014}\\

 \hline
\textbf{Calcium} &   &  \\ 
\ch{Ca^{2+}} influx &0.6 mT&Fanelli et al. (1999) \cite{FANELLI1999}\\
myosin phosphorylation in a cell-free preparation (\ch{Ca^{2+}}-dependent) &0.2 mT&Markov \& Pilla (1997)\cite{Markov1997}\\

 \ch{Ca^{2+}} concentration / morphology in cell lines &6 mT&Tenuzzo et al. (2006)\cite{Tenuzzo2006}\\

\ch{Ca^{2+}} concentration in \textit{in vitro} aged human lymphocytes &6 mT&Tenuzzo et al. (2009)\cite{Tenuzzo2009}\\

cell shape, cell surface, sugar residues, cytoskeleton, and apoptosis &6 mT&Chionna et al. (2005)\cite{Chionna2005}\\

\hline
\textbf{Neurons and brain}  &   &  \\ 
blocked sensory neuron action potentials in the somata of adult mouse  &10 mT & McLean et al. (1995) \cite{McLean1995}\\ 
 symptomatic diabetic neuropathy &50 mT& Weintraub et al (2003)\cite{Weintraub2003}\\

 \hline
 
\textbf{ROS}  &   &  \\ 
increased intercellular ROS in human neuroblastoma cells &2.2 mT& Calabr et al. (2013)\cite{Calabr2013}\\

increased intercellular ROS in human neuroblastoma cells & 31.7–232 mT& Vergallo et al. (2014)\cite{Vergallo2014}\\

increased \ch{H2O2} level in embryoid bodies&1-10 mT& Bekhite et al. (2010)\cite{Bekhite2010}\\

 ROS increase in mouse cardiac progenitor cells&0.2–5 mT &Bekhite et al. (2013) \cite{Bekhite2013}\\
 
elevated \ch{H2O2} in diploid embryonic lung fibroblast cell&230–250 mT &Sullivan et al. (2010) \cite{Sullivan2010}\\ 

increase of \ch{H2O2} in the human fibrosarcoma cancer cell&45-60 $\mu$T&Marino\&Castello (2011)\cite{Martino2011}\\

increased \ch{H2O2} production of human peripheral blood neutrophils&60 mT&Poniedziaek et al. (2013) \cite{Poniedziaek2013}\\

ROS levels in cancer cells& 10 mT& Verdon (2018) \cite{HajipourVerdom2018}\\

type 2 diabetes via regulating cellular ROS &3 mT& Carter et al. (2020 \cite{Carter2020},2021)\cite{Yu2021}\\ 

ROS changes in stem cell-mediated growth &$<$1 mT& Huizen et al. (2019) \cite{VanHuizen2019}\\

mitochondrial electron transport chain activity&0-1.93 mT&Sheu et al (2022) \cite{Sheu2022}\\

 \hline
\textbf{Others}  &   &  \\ 
flavin adenine dinucleotide photochemistry&$<$20 mT&Antill et al. (2018) \cite{Antill2018}\\

enzymatic ATP production &80 mT&Buchachenko et al. (2008) \cite{Buchachenko2008}\\
 
chlorophyll fluorescence/nutrient content of \textit{Hordeum vulgare} L.&20/42/125/250 mT&Ercan et al. (2022) \cite{Ercan2022}\\ 

antioxidant defense system of plant cells& 10/30 mT&Sahebjamei et al. (2006)\cite{Sahebjamei2006}\\ 

enhance the killing effect of adriamycin on K562 cells. &8.8 mT&Hao et al.  \cite{Hao2010}\\

regeneration and plant growth of shoot tips &2.9–4.6 mT& Atak et al. (2007)\cite{Atak2007}\\

accelerated loss of integrity of plasma membrane during apoptosis &6 mT&Teodori et al. (2002)\cite{Teodori2002}\\

 macrophagic differentiation in human pro-monocytic U937 cells &6 mT&Pagliara et al. (2009) \cite{Pagliara2009}\\

cell proliferation and cell death balance &0.5 mT&Buemi et al. (2001) \cite{Buemi2001}\\

growth and sporulation of phytopathogenic microscopic fungi  &1 mT&Nagy et al. (2004)\cite{Nagy2004}\\

  \hline
\end{tabular}
\end{adjustbox}
\caption{Static magnetic field effects on different biological functions.} 
\label{tab:SMF}
\end{table}

\subsection{Hypomagnetic field} \label{sec:HMFE}
Earth's geomagnetic field, ranging from $\sim$24 to $\sim$66 $\mu$T depending on latitude \cite{Alken2021}, can have critical roles in numerous biological processes. Shielding the geomagnetic field, called hypomagnetic field, is known to cause biological effects \cite{Belyavskaya2004,Maffei2014,Binhi2017,Zhang2020,Zhang2021rev,Xue2021,TeixeiradaSilva2015,Tsetlin2016}. 

It has also been suggested that the apparent cyclic of mass extinction on Earth \cite{Raup1984} may be related to the geomagnetic field fluctuation \cite{LIPOWSKI2006}. Decades ago, first studies on the effects of hypomagnetic field on humans were conducted, motivated by the concerns around the health of astronauts in outer space \cite{becker1963relationship,Beischer1971,beischer1967exposure,Dubrov1978}. These studies concluded that exposure to hypomagnetic fields had adverse effects on human health. Besides hypomagnetic field effects on animal and human cells and tissues, deprivation in geomagnetic field can influence the development of plants as well \cite{TeixeiradaSilva2015,Tsetlin2016}. The geomagnetic field seems to play essential roles in living organisms, and diminishing or disappearing of it could result in adverse consequences.

It was shown that exposure to hypomagnetic fields decreased the size and number of \textit{Staphylococcus aureus} Rosenbach \cite{rosenbach1884mikro}. Exposure to hypomagnetic fields can also influence early developmental processes of newt (\textit{Cynops pyrrhogaster}) \cite{Asashima1991}, early embryogenesis \cite{osipenko2008influence,Osipenko2008}, development of \textit{Xenopus} \cite{Mo2011xe}, cryptochrome-related hypocotyl growth and flowering of \textit{Arabidopsis} \cite{Xu2012,Xu2013}, development and reproduction of brown planthopper \cite{Wan2014}, mortality \cite{Erdmann2021tardigrades} and anhydrobiotic abilities \cite{Erdmann2017} in tardigrades.

It was observed that the circadian clock in fiddler crabs and other organisms \cite{Brown1960}, including human \cite{wever1970effects}, and birds \cite{BLISS1976} can be influenced by exposure to hypomagnetic fields. 

Zhang et al. showed that long-term exposure to hypomagnetic fields adversely influenced adult hippocampal neurogenesis in mice \cite{Zhang2021b}. They further observed that these effects were accompanied by reductions in ROS levels. Moreover, Wang et al. observed that exposure to hypomagnetic fields (10–100 nT) caused disorders in tubulin self-assembly \cite{Wang2008}. They show that the absorbance for monitoring tubulin self-assembly was altered by exposure to hypomagnetic fields. We discuss both these observations from the perspective of the radical pair mechanism in the following (See Sections \ref{sec:RPM-MT} and \ref{sec:RPM-NG}). Furthermore, Baek et al. reported that exposure to hypomagnetic fields influenced DNA methylation \textit{in vitro} in mouse embryonic stem cell (ESC) culture \cite{Baek2019}. Upon exposure to a hypomagnetic field ESCs morphology remained undifferentiated while under exposure to the geomagnetic field, ESCs exhibited differentiation. Moreover, Ikenaga and co-workers reported that genetic mutation in \textit{Drosophila} during space flight \cite{Ikenaga1997}. Further, Martino and co-workers reported that reducing the geomagnetic field to 6-13 $\mu$T resulted in significantly altered cell cycle rates for multiple cancer-derived cell lines \cite{Martino2010}. Belyavskaya observed that hypomagnetic conditions included reduction of the meristem, disruption of protein synthesis and accumulation of lipids, reduction of the organelle’s growth, the amount of phytoferritin in plastids and crista in mitochondria \cite{Belyavskaya2001}. Further, the effects of zero magnetic field on human VH-10 fibroblasts and lymphocytes were observed by Belyaev et al. \cite{Belyaev1997}. They concluded that exposure to hypomagnetic fields caused hypercondensation and decondensation of chromatin. Studies conducted by NASA revealed that exposure to hypomagnetic fields decreased enzyme activity in cells obtained from mice \cite{conley1970review}.   
 
Yan et al. show that reducing the magnetic field to $<$0.5 $\mu$T significantly lengthened larval and pupal development durations, increased male longevity, and reduced pupal weight, female reproduction, and the relative expression level of the vitellogenin (Vg) gene in \textit{Mythimna separata} \cite{Yan2021}. In addition, they observed that exposure to the hypomagnetic field had adverse effect on the mating ratio of \textit{M. separata} adults. They further reported that the moths in the hypomagnetic conditions had less flight activity late in the night compared to the control group. They suggest that the latter may be related to the circadian rhythm of \textit{M. separata}.

Sarimov et al. reported that hypomagnetic conditions influence human cognitive processes \cite{Sarimov2008}. They concluded that exposure to hypomagnetic fields resulted in an increased number of errors and extension of the time required to complete the tasks compared to normal conditions. \par

Wang and co-workers showed that exposure to hypomagnetic fields induced cell proliferation of SH-SY5Y cells in a glucose-dependent manner \cite{Wang2022}. They suggested that lactate dehydrogenase was a direct response to cell proliferation under hypomagnetic conditions. The authors further proposed that the up-regulation of anaerobic glycolysis and repression of oxidative stress shifted cellular metabolism more towards the Warburg effect commonly observed in cancer metabolism. Table \ref{tab:HMF} summarizes hypomagnetic field effects observed on various physiological functions.
 
\begin{table}[H]
\centering
\begin{adjustbox}{width=1\textwidth}
\begin{tabular}{lr}
  \hline
\textbf{System} &\textbf{References} \\

\hline
\textbf{Development}&\\

decrease in size and number of \textit{Staphylococcus aureus}&  Rosenbach  (1884) \cite{rosenbach1884mikro}\\ 

changes of tinctorial, morphological, cultural, and biochemical properties in bacteria & Eerkin et al. (1976)\cite{Verkin1976}\\

newt (\textit{Cynops pyrrhogaster}) - early developmental processes  & Asashima et al.(1991) \cite{Asashima1991} \\  
    
inhibition of early embryogenesis & Osipenko (2008) \cite{osipenko2008influence,Osipenko2008}\\

\textit{Xenopus} embryos- development &  Mo et al. (2011) \cite{Mo2011xe} \\  

\textit{Arabidopsis}-  cryptochrome-related hypocotyl growth and flowering &  Xu et al. (2012)\cite{Xu2012,Xu2013} \\

brown planthopper  -  development and reproduction &  Wan et al. (2014)\cite{Wan2014} \\ 
    
increased mortality in tardigrades & Erdmann et al. (2021) \cite{Erdmann2021tardigrades}\\

inhibition of anhydrobiotic abilities in tardigrades  & Erdmann et al. (2017)\cite{Erdmann2017}\\

developmental and behavioral effects in moths & Yan et al. (2021) \cite{Yan2021}\\

cell proliferation in SH-SY5Y cells, ROS implicated & Wang et al. (2022)\cite{Wang2022}\\

\hline
\textbf{Circadian system}&\\

fiddler crabs and other organisms- circadian clock & Brown (1960)\cite{Brown1960} \\
 
human -circadian rhythms& Waver et al. (1970) \cite{wever1970effects}\\

bird -circadian clock  & Bliss \& Heppner (1976) \cite{BLISS1976} \\
 
mice - circadian rhythm/ increases algesia & Mo et al. (2015) \cite{Mo2015}\\

\hline
\textbf{Neurons and brain}&\\

 inhibition of stress-induced analgesia in male mice& Seppia et al. (2000) \cite{DelSeppia2000}\\  
  
hamster -  GABA in cerebellum and basilar nucleus &  Junfeng, L.et al. (2001)\cite{junfeng2001effect} \\

mice - amnesia &  Choleris et al. (2002)\cite{Choleris2002} \\ 
   
chick -long-term memory & Wang et al. (2003) \cite{wang2003long}\\
  
impairment in learning abilities and memory of adult male mice& Wang et al. (2003) \cite{wang2003taste}  \\

\textit{Drosophila} - amnesia  & Zhang et al. (2004) \cite{Zhang2004}\\

mice-analgesia&Prato et al. (2005) \cite{Prato2005}\\
   
golden hamster-  noradrenergic activities in the brainstem &  Zhang et al. (2007)\cite{Zhang2007} \\ 

human cognitive processes & Sarimov et al. (2008) \cite{Sarimov2008}\\ 

purified tubulin from calf brain-  assembly & Wang et al. (2008) \cite{Wang2008}\\

chickens needed additional noradrenaline for memory consolidation & Xiao et al. (2009) \cite{Xiao2009}\\
   
human-  cognitive processes &  Binhi \& Sarimov (2009)\cite{BINHI2009} \\ 
human neuroblastoma cell - proliferation & Mo et al. (2013)\cite{Mo2013} \\ 
 
human neuroblastoma cells  -  actin assembly and inhibits cell motility&    Mo et al. (2016)\cite{Mo2016}\\ 

human neuroblastoma cell -\ch{H2O2} production &Zhang et al. (2017) \cite{Zhang2017}\\
  
anxiety in adult male mice &Ding et al. (2018) \cite{Ding2018}\\

mouse  - proliferation of mouse neural progenitor and stem cells & Fu et la. (2016) \cite{Fu2016sc}\\

\hline
\textbf{DNA}&\\

genetic mutations in \textit{Drosophila} during space flight& Ikenaga et al. (1997) \cite{Ikenaga1997}\\     
     
mouse embryonic stem cells (ESCs) culture-  DNA methylation & Baek et al. (2019) \cite{Baek2019} \\ 

human bronchial epithelial cells -DNA repair process & Xue et al. (2020) \cite{Xue2020}\\

\hline 
\textbf{Others} &\\

decreased enzyme activity in cells obtained from mice& Conley (1970)\cite{conley1970review}\\
 
\ch{Ca^{2+}} balance in meristem cell of pea roots& Belyavskaya (2001) \cite{Belyavskaya2001}\\

ability to change color in \textit{Xenopus laevis} &Leucht (1987)\cite{Leucht1987}\\   

chromatin hypercondensation/decondensation in human fibroblasts/lymphocytes & Belyaev et al. (1997) \cite{Belyaev1997} \\

increased protoplasts fusion & Nedukha et al. (2007)\cite{nedukha2007influence}\\

decreasing certain elements in rats' hair & Tombarkiewicz (2008) \cite{Tombarkiewicz2008}\\

cancer-derived cell lines - cell cycle rates   & Martino et al. (2010) \cite{Martino2010}\\

human fibrosarcoma cancer cells - \ch{H2O2} production & Marino et al. (2012) \cite{Martino2012} \\

mouse primary skeletal muscle cell- ROS levels &  Fu et la. (2016) \cite{Fu2016}\\

invertebrates and fish -calcium-dependent proteases &   Kantserova et al. (2017) \cite{Kantserova2017}\\

 \hline
\end{tabular}
\end{adjustbox}
\caption{Hypomagnetic field effects on different biological functions.} 
\label{tab:HMF}
\end{table}

\subsection{Oscillating magnetic field} \label{sec:OMF}

\subsubsection{Low-frequency} \label{sec:LFOMF}
The effects of oscillating magnetic fields on biological functions are abundant \cite{Lai2021,Klimek2021,Karimi2020,Pall2013,Riancho2020,Chervyakov2015,Walleczek1992,Funk2021,Moretti2022}, and are often correlated with modulation of ROS levels \cite{Schuermann2021,Nazrolu2012,Simko2007}. In this section, we review several studies on extremely low-frequency ($<$ 3 kHz) magnetic fields on various biological functions.

Sherrard and co-workers showed that exposure of the cerebellum to low-intensity repetitive transcranial magnetic stimulation (LI-rTMS) (10 mT) modulated behaviour and Purkinje cell morphology \cite{Morellini2014,Dufor2019}. Recently, the same group reported that LI-rTMS (2 mT) induced axon growth and synapse formation providing olivocerebellar reinnervation in the cerebellum \cite{Lohof2022}. The authors concluded that cryptochrome was required for the magnetosensitivity of the neurons, which was consistent with ROS production by activated cryptochrome \cite{Sherrard2018}. In a recent study, the team showed that LI-rTMS (10 mT and 10 Hz) evoked neuronal firing during the stimulation period and induced durable attenuation of synaptic activity and spontaneous firing in cortical neurons of rats \textit{in vivo} \cite{Boyer2022}. \par

Contalbrigo et al. showed that magnetic fields ($<$ 1 mT, 50 Hz) influenced some haematochemical parameters of circadian rhythms in Sprague-Dawley rats \cite{CONTALBRIGO2009}. Further, Fedele et al. reported that a 300 $\mu$T magnetic field (3-50 Hz) induced changes in two locomotor phenotypes, circadian period and activity levels via modulating cryptochrome in \textit{Drosophila} \cite{Fedele2014}. Moreover, it has been shown that exposure to a magnetic field of a 0.1 mT and 50 Hz alters clock gene expressions \cite{Manzella2015}. \par

Manikonda et al. applied magnetic fields (50 and 100 $\mu$T, 50 Hz) to cerebellum, hippocampus, and cortex of rat's brain. They observed that \ch{H2O2} increased in the descending order of cerebellum, hippocampus, and cortex. In that work, 100 $\mu$T induced more oxidative stress compared to 50 $\mu$T \cite{Manikonda2014}. Furthermore, Özgün et al. reported that exposure to a magnetic field (1 mT, 50 Hz) \textit{in vitro} induced human neuronal differentiation through N-methyl-d-aspartate (NMDA) receptor activation \cite{zgn2019}. They observed that magnetic field enhanced intracellular \ch{Ca^{2+}} levels. The authors concluded that NMDA receptors are essential for magnetosensitivity in such phenomena. It is also known that a combination of static (27-37 $\mu$T) and time varying (13/114 $\mu$T, 7/72 Hz) magnetic fields directly interact with the \ch{Ca^{2+}} channel protein in the cell membrane \cite{BaurusKoch2003}. It has also been reported that exposure to $>$5 mT (50 Hz) magnetic fields may promote X-ray-induced mutations in hamster ovary K1 cells \cite{MIYAKOSHI1999}. Koyama et al. showed that exposure to a magnetic field of 5 mT (60 Hz) promoted damage induced by \ch{H2O2}, resulting in an increase in the number of mutations in plasmids in Escherichia coli \cite{Koyama2004}. Studies of extremely low-frequency magnetic field effects ($<$ 1000 Hz) on various biological functions are shown in Tables \ref{tab:ELFMF1} and \ref{tab:ELFMF2}.

\begin{table}[ht!]
\centering
\begin{adjustbox}{width=1\textwidth}
\small
\begin{tabular}{lcr}
  \hline
\textbf{System} & \textbf{Magnetic field and frequency}  &\textbf{References} \\ 
  \hline
\textbf{Memory} &&\\

rat-acquisition and maintenance of memory&2 mT, 50 Hz& Liu et al. (2008)\cite{Liu2008}\\

rat-memory and corticosterone level&0.2 mT, 50 Hz& Mostafa et al. (2002)\cite{Mostafa2002}\\
 
spatial recognition memory in mice  &0.6/0.9/1.1/2 mT, 25/50 Hz& Fu et al. (2008) \cite{Fu_2008}\\

spatial memory disorder/ hippocampal damage in Alzheimer’s disease rat model&400 $\mu$T, 50 Hz& Liu et al. (2015) \cite{Liu2015mem}\\

recognition memory task/hippocampal spine density in mice & 1 mT, 50 Hz& Zhao et al. (2015) \cite{Zhao2015}\\

human hippocampal slices -semantic memory&1 $\mu$T, 5 min on/5 min off& Richards et al. (1996)\cite{Richards1996}\\

\hline

\textbf{Stress}&&\\

behavior/ anxiety in rats &520 $\mu$T, 50 Hz&Balassa et al. (2009) \cite{Balassa_2009}\\

benzodiazepine system in hyperalgesia in rats &0.5/1/2 mT, 60 Hz& Jeong et al. (2005)\cite{Jeong_2005}\\

anxiogenic effect in adult rats&2 mT, 50 Hz& Liu et al. (2008)\cite{Liu_2008}\\

anxiety level and spatial memory of adult rats  &2 mT, 50 Hz& He et al. (2011)\cite{he2011effects}\\

stress-related behavior of rats &   10 mT, 50 Hz & Korpinar et al. (2012)\cite{Korpinar_2012}\\

depression and corticosterone secretion in mice  &1.5/3 mT, 60 Hz& Kitaoka et al. (2012)\cite{Kitaoka_2012}\\

anxiety, memory and electrophysiological properties of male rats & 4 mT, $<$ 60 Hz & Rostami et al. (2016)\cite{Rostami_2016}\\ 

induction of anxiety via NMDA activation in mice& 1 mT, 50 Hz&Salunke et al. (2013) \cite{Salunke2013}\\

\hline

\textbf{Pain}&&\\

mice-pain thresholds& 2 mT, 60 Hz & Jeong et al. (2000)\cite{Jeong2000}\\
 
snail - analgesia & 141-414 $\mu$T, 30\&60 Hz& Prato et al. (2000)\cite{Prato2000}\\

human-analgesia/EEG&200 $\mu$T,$<$500 Hz&  Cook et al. (2004)\cite{Cook2004}\\

attenuate chronic neuropathic pain in rats& 1 mT, 1/10/20/40 Hz & Mert et al. (2017)\cite{Mert2017}\\

 mice -inhibition of morphine-induced analgesia& 0.15-9 mT, 0.5 Hz& Kavaliers \& Osscnkopp  (1987)\cite{Kavaliers1987}\\

\hline
\textbf{Dopamine / Serotonin / Melatonin}&&\\
rat frontal cortex -dopamine and serotonin level &1.8-3.8 mT, 10 Hz& Siero et al. (2004)\cite{Siero2004}\\

rat brain - serotonin and dopamine receptors activity & 0.5 mT,50 Hz & Janac et al. (2009) \cite{janac2009effect}\\

rat -central dopamine receptor &1.8–3.8 mT, 10 Hz& Siero2001 et al. (2001)\cite{Siero2001}\\

rat -plasma and pineal melatonin levels&1/5/50/250 $\mu$T, 50 Hz&Kato et al. (1993)\cite{Kato1993}\\

human -  melatonin concentration& 2.9 mT, 40 Hz&Karasek et al. (1998) \cite{Karasek1998}\\

\hline  

\textbf{Genetic}&&\\
rat brain cells -increases DNA strand breaks&0.5 mT, 60 Hz& Lai \& Singh (1997)\cite{Lai1997,Lai1997melat}\\ 

human HL-60 cells-steady-state levels of some mRNAs &8 $\mu$T, 60 Hz & Karabakhtsian et al. (1994) \cite{Karabakhtsian1994}\\
 
hamster ovary K1cells-promotion in X-ray-induced mutations &$>$5 mT, 50 Hz&  Miyakoshi et al. (1999)\cite{MIYAKOSHI1999}\\

HL60 cells - CREB DNA binding activation& 0.1 mT, 50 Hz &Zhou et al. (2002) \cite{Zhou2002}\\

plasmids in Escherichia coli-increase in the number of mutations &5 mT, 60 Hz &Komaya et al. (2004) \cite{Koyama2004}\\

genetic analysis of circadian responses in \textit{Drosophila} &300 $\mu$T, 3-50 Hz& Fedele et al. (2014) \cite{Fedele2014}\\

epigenetic modulation of adult hippocampal neurogenesis in mice & 1 mT, 50 Hz& Leone et al. (2014) \cite{Leone2014}\\

circadian gene expression in human fibroblast cell  & 0.1 mT, 50 Hz & Manzella et al. (2015)\cite{Manzella2015}\\

epigenetic modulation in human neuroblastoma cells &1 mT, 50 Hz & Consales et al. (2017)\cite{Consales2017}\\

\hline  

\textbf{Calcium}&&\\  
  
lymphocyte - calcium signal transduction  &42.1 $\mu$T, 16 Hz &Yost \& Liburdy (1992)\cite{Yost1992} \\
  
T-cell- intracellular calcium oscillations   & 0.1 mT, 50 Hz& Lindströum et al. (1993) \cite{Lindstrum1993}\\   

rat pituitary cells -\ch{Ca^{2+}} influx &50 $\mu$T, 50 Hz &Barbier et al. (1996) \cite{Barbier1996}\\

\ch{Ca^{2+}} channel protein in the cell membrane &  13/114 $\mu$T, 7/72 Hz  & BaurusKoch et al. (2003) \cite{BaurusKoch2003}\\

human skin fibroblast populations  - intracellular calcium Oscillations  & 8 mT, 20 Hz& Löschinger et al. (1999) \cite{Lschinger1999}\\
  
osteoblasts cells - intracellular calcium levels & 0.8 mT, 50 Hz& Zhang et al. (2010)\cite{Zhang2010}\\

C2C12 muscle cells - calcium handling and increasing \ch{H2O2}&1 mT, 50 Hz &Morabito et al. (2010)\cite{Morabito2010}\\ 

rat ventricle cells- intracellular \ch{Ca^{2+}}  &0.2 mT, 50 Hz &Sert et al. (2011)\cite{Sert2011}\\

mesenchymal stem cells  - \ch{Ca^{2+}} intake &1 mT, 50 Hz & Özgün \& Garipcan (2021)\cite{zgn2021}\\

brain tissue - radiation-induced efflux of \ch{Ca^{2+}} ions &$\mu$T, 15/45 Hz&Blackman et al. (1985) \cite{Blackman1985}\\
  
rat hippocampus-\ch{Ca^{2+}} signaling and NMDA receptor functions&50/100 $\mu$T, $<$300 Hz& Manikonda et al. (2007)\cite{Manikonda2007}\\ 

entorhinal cortex neurons - calcium dynamics&1/3 mT, 50 Hz& Luo er al. (2014) \cite{Luo2014}\\

  \hline
\end{tabular}
\end{adjustbox}
\caption{Extremely low-frequency  ($<$ 3 kHz) magnetic field effects on memory, stress, pain, dopamine, serotonin, melatonine, genetics, and calcium flux.} 
\label{tab:ELFMF1}
\end{table}

\begin{table}[ht!]
\centering
\begin{adjustbox}{width=1\textwidth}
\small
\begin{tabular}{lcr}

\hline

\textbf{System} & \textbf{Magnetic field}  &\textbf{References} \\

\hline  

\textbf{ROS}&&\\  

ageing via ROS involvement in brain of mongolian gerbils&0.1/0.25/0.5 mT, 50 Hz& Selakovi et al. (2013)\cite{Selakovi2013}\\ 
 
hippocampus mitochondria via increasing \ch{H2O2} in mice &8 mT, 50 Hz&Duan et al. (2013)\cite{Duan2013}\\

neural differentiation/ \ch{H2O2} elevation in mesenchymal stem cells &1 mT, 50 Hz&Park et al. (2013)\cite{Park2013}\\

 \ch{H2O2} production in neuroblastoma cell  & 2$\pm$0.2 mT, 75$\pm$2 Hz & Osera et al. (2015) \cite{Osera2015}\\
 
 pro-Parkinson’s disease toxin MPP$^+$/ \ch{H2O2} increase in SH-SY5Y cells & 1 mT, 50 Hz&Benassi et al. (2015)\cite{Benassi2015}\\

rat peritoneal neutrophils -oxidative burst&0.1 mT, 60 Hz & Roy et al. (1995)\cite{Roy1995}\\

cortical synaptosomes of Wistar rats-oxidative stress&0.7 mT, 60 Hz& Túnez et al. (2006) \cite{Tnez2006}\\

 pro-oxidant effects of \ch{H2O2} in human neuroblastoma cells &2 mT, 75 Hz& Falone et al. (2016) \cite{Falone2016}\\ 
 reducing hypoxia/inflammation damage ROS-mediated in neuron-like and microglial cells &1.5$\pm$0.2 mT, 75 Hz& Vincenzi et al. (2016)\cite{Vincenzi2016}\\ 

mouse brain-antioxidant defense system&1.2 mT, 60 Hz& Lee et al. (2004) \cite{Lee2004}\\

rat-cortical neurons -redox and trophic response/ reducing ROS &1 mT, 50 Hz& DiLoreto et al. (2009)\cite{DiLoreto2009}\\

human monocytes-cell activating capacity/ROS modulation&1 mT, 50 Hz& Lupke et al. (2004) \cite{Lupke2004}\\

HL-60 leukemia cells- proliferation / DNA damage implicating ROS   &1 mT, 50 Hz& Wolf et al. (2005)\cite{Wolf2005}\\

human monocytes-alteration of 986 genes/ modulating ROS &1 mT, 50 Hz& Lupke et al. (2006) \cite{Lupke2006} \\

prostate cancer cells -apoptosis  through ROS&0.2 mT, 60 Hz&Koh et al. (2008) \cite{Koh2008}\\
 
K562 cells -\ch{O2^{.-}} formation and HSP70 induction  & 0.025–0.1 mT, 50 Hz& Mannerling et al. (2010)\cite{Mannerling2010}\\

K562 Cells -differentiation via increasing \ch{O2^{.-}} production & 5 mT, 50 Hz & AySe et al. (2010)\cite{Aye2010}\\

K562  leukemia cell -number of apoptotic cells via increasing \ch{O2^{.-}} production & 1 mT, 50 Hz& Garip \& Akan (2010)\cite{Garip2010}\\

PC12 cells -\ch{H2O2} increase&1 mT, 50 Hz&Morabito et al. (2010) \cite{Morabito2010b}\\

carcinoma cells  -  cisplatin via increasing \ch{H2O2} &1 mT, 50 Hz&Bułdak et al. (2012) \cite{Budak2012}\\

human carcinoma cells   -morphology and biochemistry implicating ROS&0.1 mT, 100\&217 Hz& Sadeghipour et al. (2012)\cite{Sadeghipour2012}\\ 
 
rats-  DNA strand breaks in brain cells by modulating ROS&0.1–0.5 mT, 60 Hz &Lai \& Singh(2004)\cite{Lai2004}\\

cardiomyocytes-injury treatment implicating ROS&4.5 mT, 15 Hz&Ma et al. (2013)\cite{Ma2013}\\ 

 genomic instability/oxidative processes in human neuroblastoma cells& 100$\mu$T, 50 Hz& Luukkonen et al. (2014)\cite{Luukkonen2014}\\

expression of NOS and \ch{O2^{.-}} in human SH-SY5Y cells&1 mT, 50 Hz & Reale et al. (2014)\cite{Reale2014}\\

ROS-related autophagy in mouse embryonic fibroblasts&2 mT, 50 Hz &Chen et al. (2014)\cite{Chen2014}\\

healing via reducing ROS production in artificial skin wounds&$<$ 40 $\mu$T, 100 Hz&Ferroni et al. (2015)\cite{Ferroni2015}\\

apoptosis via oxidative stress in human osteosarcoma cells &1mT, 50 Hz& Yang et al. (2015) \cite{yang2015extremely}\\

increase \ch{O2^{.-}} in erythro-leukemic cells &1 mT, 50 Hz&Patruno et al. (2015) \cite{Patruno2015}\\

Genomic instability/ \ch{H2O2} increase in SH-SY5Y cells  & 100 $\mu$T, 50 Hz & Kesari et al. (2015)\cite{Kesari2015}\\

Nox-produced ROS in hAECs & 0.4 mT, 50 Hz & Feng et al. (2016) \cite{Feng2016}\\  

mitochondrial permeability via increasing \ch{H2O2} in human aortic endothelial cells  &0.4 mT, 50 Hz &Feng et al. (2016) \cite{Feng2016c}\\

apoptotic via mitochondrial \ch{O2^{.-}} release in human aortic endothelial cells  &0.4 mT, 50 Hz& Feng et al. (2016)\cite{Feng2016b}\\  

antioxidant activity implicating \ch{H2O2} in human keratinocyte cells &25-200 $\mu$T, 1-50 Hz &Calcabrini et al. (2016) \cite{Calcabrini2016}\\

antioxidative defense mechanisms via ROS in human osteoblasts & 2-282 $\mu$T, 16 Hz, &Ehnert et al. (2017)\cite{Ehnert2017}\\ 

astrocytic differentiation implicating ROS in human bone  stem cells &1 mT, 50 Hz &Jeong et al. (2017)\cite{Jeong2017}\\ 

reduce mitochondrial \ch{O2^{.-}} production in human neuroblastoma cells& 100 $\mu$T, 50 Hz & Höytö et al. (2017)\cite{Hyt2017}\\ 

ROS production  in human cryptochrome& 1.8 mT, $<$100 Hz & Sherrard et al. (2018) \cite{Sherrard2018}\\ 

proliferation by decreasing intracellular ROS levels in human cells & 10 mT, 60 Hz & Song et al. (2018)\cite{Song2018}\\

cytotoxic effect in  by raising intracellular ROS in human GBM cells & 1–58 mT, 350 Hz&Helekar et al. (2021)\cite{Helekar2021}\\

\hline

\end{tabular}
\end{adjustbox}
\caption{Extremely low-frequency ($<$ 3 kHz) magnetic field effects on reactive oxygen species (ROS) levels.} 
\label{tab:ELFMF2}
\end{table}

\begin{table}[ht!]
\centering
\begin{adjustbox}{width=1\textwidth}
\small
\begin{tabular}{lcr}
  \hline
\textbf{System} & M\textbf{agnetic field}  &\textbf{References} \\ 

\hline
\textbf{Others}&&\\

neuroendocrine cell-proliferation and death&$<$1 mT, 50 Hz& Grassi et al. (2004) \cite{Grassi2004}\\

cortices of mice-neuronal differentiation of neural stem/progenitor cells&1 mT, 50 Hz&Piacentini et al. (2008) \cite{Piacentini2008}\\

hippocampal slices - excitability in hippocampal neurons&15 mT, 0.16 Hz& Ahmed \& Wieraszko (2008) \cite{Ahmed2008}\\

human -EEG alpha activity& 200 $\mu$T, 300 Hz& Cook et al. (2009) \cite{Cook2006,Cook2009}\\

rat -neuroprotective effects &0.1/0.3/0.5 mT, 15 Hz& Yang et al. (2012)\cite{Yang2012}\\

rat  -neuroprotective effects on Huntington's disease&0.7 mT, 60 Hz& Tasset et al. (2012) \cite{Tasset2012}\\

synaptic efficacy in rat brain slices &0.5/3 mT, 50 Hz& Balassa et al. (2013)\cite{Balassa2013}\\

global cerebral ischemia / pituitary ACTH and TSH cells in gerbils&0.5 mT, 50 Hz& Balind et al. (2019) \cite{Rau_Balind_2019}\\ 

neurotrophic factor expression in rat dorsal root ganglion neurons&1 mT, 50 Hz& Li et al. (2014) \cite{Li2014}\\

 visual cortical circuit topography and BDNF in mice &$\sim$10 mT, $<$10 Hz  &  Makowiecki et al. (2014)  \cite{Makowiecki2014}\\

 hippocampal long-term potentiation in rat   &100 $\mu$T, 50 Hz& Komaki et al. (2014) \cite{Komaki2014}\\
 
 neuronal GABAA current in rat cerebellar granule neurons & 1 mT, 50 Hz&Yang et al. (2015)\cite{Yang2015}\\

central nervous regeneration in planarian Girardia sinensis &200 mT, 60 Hz& Chen et al. (2016) \cite{Chen2016}\\

neuronal differentiation and neurite outgrowth in embryonic neural stem cells&1 mT, 50 Hz& Ma et al. (2016) \cite{Ma2016}
\\
 
synaptic transmission and plasticity in mammalian central nervous synapse&1mT, 50 Hz&Sun et al. (2016)\cite{Sun2016}\\
  
human - pineal gland function & $<$$\mu$T, 60 Hz& Wilson et al. (1990)\cite{Wilson1990}\\
 
rat - electrically kindled seizures& 0.1 mT, 60 Hz & Ossenkopp \& Cain (1988)\cite{Ossenkopp1988} \\ 

rat -central cholinergic systems   &  1 mT, 60 Hz & Lai et al.  (1993)\cite{Lai1993}\\

deer mice -spatial learning&0.1 mT, 60 Hz& Kavaliers et al. (1996) \cite{Kavaliers1996}\\

T cell receptor - signalling pathway & 0.15 mT, 50 Hz &  Lindstrm et al.(1998)\cite{Lindstrm1998}\\

enhances locomotor activity via activation of dopamine D1-like receptors in mice &0.3/2.4 mT, 60 Hz& Shin et al. (2007)\cite{Shin_2007}\\

rat pituitary ACTH cells &0.5 mT, 50 Hz& Balind et al. (2014)\cite{Rau_Balind_2014}\\ 
 
actin cytoskeleton reorganization in human amniotic cells &0.4 mT, 50 Hz& Wu et al. (2014) \cite{Wu2014}\\

reduces hypoxia and inflammation in damage microglial cells &1.5 mT, 50 Hz & Vincenzi et al. (2016)\cite{Vincenzi2016}\\

pluripotency and neuronal differentiation in mesenchymal stem cells &20 mT, 50 Hz & Haghighat et al. (2017)\cite{Haghighat2017}\\

proliferation and differentiation in osteoblast cells & 5 mT, 15 Hz & Tong et al. (2017)\cite{Tong2017}\\ 

reduced hyper-inflammation triggered by COVID-19 in human & 10 mT, 300 Hz& Pooam et al. (2021)\cite{Pooam2021}\\

proliferation and regeneration in planarian Schmidtea mediterranea& 74 $\mu$T, 30 Hz & Ermakov et al. (2022)\cite{Ermakov2022}\\ 

  \hline
\end{tabular}
\end{adjustbox}
\caption{Extremely low-frequency ($<$ 3 kHz) magnetic field effects on different biological functions.} 
\label{tab:ELFMF3}
\end{table}

\subsubsection{Medium/High-frequency} \label{sec:HFOMF}

In this section, we review several studies on medium/high-frequency ($> 3kHz$) magnetic field effects on various physiological functions. Usselman et al. reported that oscillating magnetic fields at Zeeman resonance  (1.4 MHz and 50 $\mu$T) influenced relative yields of cellular \ch{O2^{.-}} and \ch{H2O2} products in human umbilical vein endothelial cells \cite{Usselman2016}. Considering a radical pair in [\ch{FADH^{.}}..\ch{O2^{.-}}] form, the authors suggested that coherent electron spin dynamics may explain their observation. Moreover, Friedman et al. observed that a 875 MHz magnetic field increased ROS production, which was mediated by membrane-associated Nox in HeLa cells and rat \cite{Friedman2007}. Castello and colleagues showed that exposure of fibrosarcoma HT1080 cells to weak radio frequency (5/10 MHz) combined with a 45 $\mu$T static magnetic field modulated the number of cells and significantly increased \ch{H2O2} production \cite{Castello2014}. Martino and Castello showed that exposure of cultured yeast and isolated mitochondria to magnetic fields (150 $\mu$T; 45 $\mu$T and a parallel 10 MHz RF; 45 $\mu$T and a perpendicular 10 MHz RF) modulated the production of extracellular, intracellular, and mitochondrial \ch{O2^{.-}} and \ch{H2O2} \cite{Martino2013}. They concluded that complex I of the electron transport chain is involved in the  \ch{H2O2} production. Table \ref{tab:MHFMF} summarizes a few medium/high-frequency magnetic field effects observed in various experiments.

\begin{table}[ht!]
\centering
\begin{adjustbox}{width=1\textwidth}
\begin{tabular}{lcr}

\hline

\textbf{System} &\textbf{ Magnetic field and frequency}  & \textbf{References} \\

\hline

ROS production and DNA damage in human SH-SY5Y neuroblastoma cells&872 MHz&  Luukkonen et al. (2009) \cite{Luukkonen2009}\\  

ROS level in human ejaculated semen &870 MHz& Agarwal et al (2009)  \cite{Agarwal2009}\\

ROS Production and DNA Damage in human spermatozoa &1.8 GHz& Iuliis et al (2009) \cite{DeIuliis2009}\\

ROS levels and DNA fragmentation  in astrocytes &900 MHz& Campisi et al. (2010)  \cite{Campisi2010}\\

ROS Formation and apoptosis in human peripheral blood mononuclear cell &900 MHz& Lu et al. (2012) \cite{Lu2012}\\ 

ROS elevation in \textit{Drosophila}&1.88–1.90 GHz& Manta et al. (2013) \cite{Manta2013}\\ 
  
ROS modulation in rat pulmonary arterial smooth muscle cells &7 MHz&Usselman et al. (2014)\cite{Usselman2014} \\ 

bioluminescence and oxidative response in HEK cells&940 MHz&Sefidbakht et al. (2014)\cite{Sefidbakht2014}\\

electrical network activity in brain tissue &  $<$150 MHz& Gramowski-Vo{\ss} et al. (2015) \cite{GramowskiVo2015}\\

ROS production in human umbilical vein endothelial cells&50 $\mu$T, 1.4 MHz&Usselman et al. (2016)\cite{Usselman2016}\\ 

insect circadian clock & 420 $\mu$T, RF& Bartos et al. (2019) \cite{Bartos2019}\\

tinnitus, migraine and non-specific in human & 100 KHz to 300 GHz & Röösli et al. (2021) \cite{Rsli2021} \\
 
magnetic compass orientation in night-migratory songbird  & 75–85 MHz & Leberecht et al. (2022)\cite{Leberecht2022}\\

\hline

\end{tabular}
\end{adjustbox}
\caption{Medium/High-frequency ($>$ 3 kHz) magnetic field effects on biological functions.} 
\label{tab:MHFMF}
\end{table}

\subsection{Isotope effects}\label{sec:MIE}

Atomic nuclei contain protons and neutrons. The number of protons determines the element (e.g. carbon, oxygen, etc.), and the number of neutrons determines the isotope of the desired element. Some isotopes are stable, i.e. they preserve the number of protons and neutrons during chemical reactions. It has been shown that using different isotopes of the element in certain chemical reactions results in different outcomes. Such observations have been seen in many chemical reactions \cite{Bigeleisen1965,Zeldovich1988,Wolfsberg2009,faure1977principles,hoefs2009stable,fry2006stable,van2011isotope,Buchachenko2001} including biological processes \cite{cook1991enzyme,Grissom1995,kohen2005isotope,buchachenko2009magnetic,Buchachenko2012,Koltover2021}. Inheriting quantum properties, not only do different isotopes of an element have different masses, but they can also have different spins. For that reason, isotope effects in (bio)chemical reactions can be regarded from two distinct points of view: mass-dependency and spin-dependency. Thiemens et al. observed mass-independent isotope effects as a deviation of isotopic distribution in reaction
products \cite{Thiemens1983,Thiemens1999,Thiemens2001,Thiemens2006,Thiemens2012}. Furthermore, in 1976 Buchachenko and colleagues by applying magnetic fields detected the first mass-independent isotope effect, which chemically discriminated isotopes by their nuclear spins and nuclear magnetic moments \cite{buchachenko1976isotopic}. Since then, the term "magnetic isotope effect" was dubbed for such phenomena as they are controlled by electron-nuclear hyperfine coupling in the paramagnetic species. Moreover, isotope effects have been observed for a great variety of chemical and biochemical reactions involving oxygen, silicon, sulfur, germanium, tin, mercury, magnesium, calcium, zinc, and uranium \cite{Buchachenko2012,Buchachenko2013,Buchachenko2014,LBuchachenko2014,bukhvostov2014new,Buchachenko2019,Arkhangelskaya2020,Buchachenko2020,Koltover2021,Letuta2021}. In this review, we focus on isotope effects from spin perspective, see Table \ref{tab:MIE}. 
\par

In 1986 Sechzer and co-workers reported that lithium administration results in different parenting behaviors and potentially delayed offspring development in rats \cite{Sechzer1986}. Their findings weren't quantitative; however, it was observed that different lithium isotopes exhibited different impacts. Moreover, in 2020, Ettenberg \textit{et al.} \cite{Ettenberg2020}  conducted an experiment demonstrating an isotope effect of lithium on rat's hyperactivity. Lithium has two stable isotopes, \ce{^{6}Li} and \ce{^{7}Li}, possessing different nuclear spin angular momentum, $I_6=1$ and $I_7=3/2$, respectively. In that work, the mania phase was induced by sub-anesthetic doses of ketamine. The authors reported that \ce{^{6}Li} produced a longer suppression of hyperactivity in an animal model of mania compared to \ce{^{7}Li}. We further discuss this phenomenon from the point of view of the radical pair mechanism in Section \ref{sec:RPM-Li}.  
\par
Li and co-workers reported that xenon (Xe)-induced anesthesia in mice is isotope-dependent. They used four different Xe isotopes, \ce{^{129}Xe}, \ce{^{131}Xe}, \ce{^{132}Xe}, and \ce{^{134}Xe} with nuclear spins of 1/2, 3/2, 0, and 0, respectively \cite{Li2018}. The results fell into two groups, isotopes with spin and isotopes without spin, such that isotopes of xenon with non-zero nuclear spin had lower anesthetic potency than isotopes with no nuclear spin. The results of this work are discussed from the perspective of the radical pair mechanism in Section \ref{sec:RPM-Xe}.
\par
Buchachenko et al. observed that magnesium-25 (\ch{^{25} Mg})  controlled phosphoglycerate kinase (PGK) \cite{Buchachenko2005}. \ch{^{25} Mg} has a nuclear spin of 5/2, while \ch{^{24} Mg} is spin-less. The authors reported that ATP production was more than twofold in the presence of \ch{^{25} Mg} compared to \ch{^{24} Mg}. They suggested that the nuclear spin of Mg was the key factor for such an observation. In another study, the same group reported that \ch{^{25} Mg} reduced enzymatic activity compared to \ch{^{25} Mg} in polymerases $\beta$. They concluded that DNA synthesis is magnetic-dependent \cite{Buchachenko2013d,Buchachenko2013DNA}. In the same system, they further observed that if \ch{Mg^{2+}} ion is replaced by stable isotopes of calcium ion, \ch{^{40}Ca^{2+}} and \ch{^{43}Ca^{2+}} (with nuclear spins of 0, 7/2, respectively), the enzyme catalytic reactions will be isotope-dependent, such that \ch{^{43}Ca^{2+}}promoted enzyme hyper-suppression leading to a residual synthesis of shorted DNA fragments compared to \ch{^{40}Ca^{2+}} \cite{bukhvostov201343}. They repeated the same experiment but his time instead of \ch{Mg^{2+}} ion stable isotopes of zinc, \ch{^{64}Zn^{2+}} and \ch{^{67}Zn^{2+}} (with nuclear spins of 0, 5/2, respectively) were used. The authors reported that \ch{^{67}Zn^{2+}} suppressed DNA synthesis a few times more than \ch{^{64}Zn^{2+}} \cite{Buchachenko2010a}.   
\par

\begin{table}[ht!]
\centering
\begin{adjustbox}{width=1\textwidth}
\begin{tabular}{lccr}
  \hline
\textbf{System} &  \textbf{Isotope} & \textbf{Spin}, $I$ & \textbf{References} \\ 
\hline

parenting/offspring development in rat &   \ch{^6Li}, \ch{^7Li} & 1, 3/2 & Sechzer et al. (1986) \cite{Sechzer1986} \\ 

hyperactivity in rat &\ch{^6Li}, \ch{^7Li} & 1, 3/2 & Ettenberg et al. (2020) \cite{Ettenberg2020} \\ 

anesthetic potency in mice &\ch{^{129}Xe}, \ch{^{131}Xe}, \ch{^{132}Xe}, \ch{^{134}Xe} & 1/2, 3/2, 0, 0 & Li et al. (2018) \cite{Li2018} \\ 

ATP production in purified pig skeletal muscle PGK&   \ch{^{24}Mg}, \ch{^{25}Mg}, \ch{^{26}Mg} & 0, 5/2, 0 & Buchachenko et al. (2005) \cite{Buchachenko2005}\\

DNA synthesis in HL-60 human myeloid leukemia cells    &\ch{^{64}Zn}, \ch{^{67}Zn} & 0, 5/2 & Buchachenko et al. (2010) \cite{Buchachenko2010a}\\

DNA synthesis in HL-60 human myeloid leukemia cells &\ch{^{24}Mg}, \ch{^{25}Mg}, \ch{^{26}Mg} & 0, 5/2, 0  & Buchachenko et al. (2013) \cite{Buchachenko2013d}\\

DNA synthesis in HL-60 human myeloid leukemia cells &\ch{^{40}Ca}, \ch{^{43}Ca} & 0, 7/2 & Bukhvostov et al. (2013) \cite{bukhvostov201343}\\

 \hline
\end{tabular}
\end{adjustbox}
\caption{Spin-dependent isotope effects on different biological functions.} 
\label{tab:MIE}
\end{table}

\section{The radical pair mechanism} \label{sec:RPM}

\subsection{Spin and radical pairs}\label{sec:spin-RP}
Spin is an inherently quantum property that emerges from Dirac's relativistic quantum mechanics \cite{Dirac1928,Ohanian1986}, and is described by two numbers, $S$ and $m_s$, respectively, the spin quantum number and the spin projection quantum number. Electrons, protons, and neutrons have spins of $S=\frac{1}{2}$. Having an angular momentum characteristic, spin can be coupled not only with external magnetic fields but also with other spin in its vicinity. For instance, coupling of two electrons spins, \textbf{S$_A$} and \textbf{S$_B$}, results in a total spin of \textbf{S$_T$}, which has a quantum number of either $S$ = 1 or $S$ = 0. The latter case is called a singlet state, with $m_s$ = 0, and the former is called a triplet state, with $m_s$ = 0, $\pm$1 \cite{Sakurai1995}.

\begin{ceqn}
\begin{equation}
 \ket{S} = \frac{1}{\sqrt{2}} \Big( \ket{\uparrow}_A \otimes \ket{\downarrow}_B -\ket{\downarrow}_A \otimes \ket{\uparrow}_B\Big),
 \label{eq:singlet}
\end{equation}
\end{ceqn}

\begin{ceqn}
\begin{equation}
 \ket{T_-} = \ket{\downarrow}_A \otimes \ket{\downarrow}_B,
\end{equation}
\end{ceqn}

\begin{ceqn}
\begin{equation}
 \ket{T_0} = \frac{1}{\sqrt{2}} \Big(\ket{\uparrow}_A \otimes \ket{\downarrow}_B +\ket{\downarrow}_A \otimes \ket{\uparrow}_B\Big),
\end{equation}
\end{ceqn}

\begin{ceqn}
\begin{equation}
 \ket{T_+} = \ket{\uparrow}_A \otimes \ket{\uparrow}_B,
\end{equation}
\end{ceqn}
where $\otimes$ is the tensor product.
\par
Radicals are molecules with an odd number of electrons in the outer shell \cite{salikhov1984spin,Gerson2003}. A pair of radicals can be formed by breaking a chemical bond or electron transfer between two molecules. It is important to notice that in reactions of organic molecules, spin is usually a conserved quantity, which is essential for magnetic field effect in biochemical reactions. For example, a radical pair can be created if a bond between a pair of molecules [A...D] breaks or an electron is transferred from D to A, [\ch{A^{-.}}...\ch{D^{.+}}] (D and A denote donor and acceptor molecules). A radical pair may be in a superposition of singlet and triplet states, depending on the parent molecule’s spin configuration. Assuming that the initial state of the electron pairs before separation was a singlet (triplet), the recombination of unpaired electrons can only happen if they stayed in a singlet (triplet) \cite{Hayashi2004}. 

If the radical pairs are formed in singlet (triplet) states, the initial spin density matrix reads as follows:
\begin{ceqn}
\begin{equation}
 \hat{\rho}(0) = \frac{1}{M} \hat{P}^S , 
\end{equation}
\end{ceqn}

\begin{ceqn}
\begin{equation}
\hat{P}^S= \ket{S} \otimes \bra{S} \otimes \hat{\mathbbm{1}}_{M},
\end{equation}
\end{ceqn}

\begin{ceqn}
\begin{equation}
\hat{P}^T= \big\{ \ket{T_+} \otimes \bra{T_+} + \ket{T_0} \otimes \bra{T_0} + \ket{T_-} \otimes \bra{T_-}\big\} \otimes \hat{\mathbbm{1}}_{M},
\end{equation}
\end{ceqn}

\begin{ceqn}
\begin{equation}
\hat{P}^S + \hat{P}^T= \hat{\mathbbm{1}}_{4M},
\end{equation}
\end{ceqn}

\begin{ceqn}
\begin{equation}
M =\prod_{i}^n (2 I_i +1), 
\end{equation}
\end{ceqn}
where $\hat{P}^S$ and $\hat{P}^T$ are the singlet and triplet projection operators, respectively, $M$ is the nuclear spin multiplicity, $I_i$ is the spin angular momentum of $i$-th nucleus, and $\hat{\mathbbm{1}}$ is the identity matrix. S is entangled. The T projector is not entangled, even though T0 is an entangled state. 

\subsection{Interactions}

\subsubsection{Zeeman interaction}
The interaction between the unpaired electron spins on each radical and the external magnetic field is essential for generating MFEs. This interaction is called the Zeeman effect \cite{wertz2012electron}. The nuclear spins of radical molecules also experience applied magnetic fields; however, as nuclear magnetogyric ratios are much smaller than that of the electrons, these interactions are negligible. The Zeeman interaction is defined in the following form:

\begin{ceqn}
\begin{equation}
 \hat{H}_Z = \mu_{B} \mathbf{\hat{S}}.\textbf{g}.\mathbf{B},
 \label{eq:Zeeman1}
\end{equation}
\end{ceqn}
where $\mu_{B}$, $\mathbf{\hat{S}}$, $\textbf{g}$-tensor, and $\mathbf{B}$ are the Bohr magneton, the spin operators of electron, the interaction coupling, and applied magnetic field, respectively. Here, we focus on magnetic field interactions with relatively low field strengths. In such cases, it is possible to assume that the $\textbf{g}$-tensor equals to $g_e$ of free electron, and hence,
\begin{ceqn}
\begin{equation}
 \hat{H}_Z = g_e \mu_{B} \mathbf{\hat{S}}.\mathbf{B} = - \gamma_e \mathbf{\hat{S}}.\mathbf{B},
 \label{eq:Zeeman2}
\end{equation}
\end{ceqn}
where $\gamma_e$ is the electron magnetogyric ratio.

\subsubsection{Hyperfine interaction}\label{sec:HFI}
Similar to electron-electron spin coupling, electron spins can couple to the nuclear spins, called hyperfine interactions \cite{atkins2011molecular}. This interaction consists of two contributions, isotropic and anisotropic interactions. The former is also called Fermi contact term, which results from the magnetic interaction of the electron and nuclear spins when the electron is \textit{within} the nucleus. The overall hyperfine interaction can be defined as follows:

\begin{ceqn}
\begin{equation}
 \hat{H}_{HFI} = \mathbf{\hat{S}}.\textbf{a}_i.\mathbf{\hat{I}}_i,
 \label{eq:HFI}
\end{equation}
\end{ceqn}
where $\textbf{a}_i$ and $\mathbf{\hat{I}}_i$ are the hyperfine coupling tensor and nuclear spin of $i$-th nucleus. The anisotropic components of the hyperfine interactions are only relevant when the radicals are immobilised and aligned \cite{Schulten1978}. Neglecting the anisotropic component of the hyperfine interaction, the hyperfine Hamiltonian has the following form:
\begin{ceqn}
\begin{equation}
 \hat{H}_{HFI} = a_i \mathbf{\hat{S}}.\mathbf{\hat{I}}_i,
 \label{eq:HFI1}
\end{equation}
\end{ceqn}
where $a_i$ is the isotropic hyperfine coupling constant and can be calculated as:
\begin{ceqn}
\begin{equation}
 a_i = - \frac{2}{3} g_e \gamma_e \gamma_n \mu_0 |\Psi(0)|^2,
 \label{eq:HFI2}
\end{equation}
\end{ceqn}
$\mu_0$ is the vacuum permeability, $\gamma_n$ is the nuclear magnetogyric ratio, and $|\Psi(0)|^2$ is the electron probability density at the nucleus \cite{Improta2004}.

\subsubsection{Exchange interaction}
The electrons on radicals are identical in quantum calculations. This indistinguishability of electrons on radical pairs can be introduced via the exchange interaction \cite{Illas2000}. It is generally assumed to weaken exponentially with increasing radical pair separation. The exchange interaction can prevent singlet-triplet interconversion, as discussed later. However, recent studies show that this term is negligible \cite{Nohr2017} in the magnetic field effects on pigeon cryptochrome \cite{Hochstoeger2020}.

\subsubsection{Dipolar interaction}
As spins are magnetic moments, the radical pairs also influence each other by a dipolar interaction \cite{ernst1987principles}. This interaction can suppress singlet–triplet interconversion in the radical pair dynamics. However, studies on avian magnetoreception suggest that under certain conditions exchange and dipolar interactions can be neglected \cite{Efimova2008,Hore2016,Babcock2021,Kattnig2017,Kattnig2017a}.

\subsubsection{Other contributions}
It is thought that after a first re-encounter, radicals either react or diffuse apart forever \cite{BROCKLEHURST1996}. In the context of brids' magnetoreception, for this contribution, an exponential model is used \cite{Hore2019,Hore2016}.
\par
High electron density on an atom of a radical can lead to have a higher anisotropic $g$-value compared to the case with lower electron density, called spin-orbit effect, which results in the non-radiative transition between two electronic states with different spin multiplicity (e.g. singlet and triplet)-- intersystem crossing, which can play important roles in chemical reactions \cite{hameka19671,Khudyakov1993,Li2018ISC,Marian2021}.

\subsection{Spin dynamics of radical pairs}

The sensitivity of certain reactions to weak magnetic fields relies on the oscillations between singlet and triplet states of radical pairs, also known as "quantum beats" \cite{Steiner1989}. If the radicals are spatially enough separated, having the same energies, singlet and triplet will undergo a coherent interconversion process, quantum beating. The interconversion is tuned by the magnetic fields experienced by the electrons, including Zeeman and hyperfine interactions. At low magnetic fields, the main drive for S-T interconversion is due to the hyperfine interactions. Obeying selection rules, the singlet and triplet yields will follow different chemical pathways, which depend on the timing of the coherent spin dynamics \cite{Hore2021}. These quantum beats have just recently been observed directly \cite{Mims2021}. \par

The fractional singlet yield resulting from the radical pair mechanism throughout the reaction can be normally defined by using the Liouville–von Neumann equation \cite{Timmel1998}:

\begin{ceqn}
\begin{align}
 \frac{d \hat{\rho}(t)}{dt} = -\frac{i}{\hbar}[\hat{H},\hat{\rho}(t)],
 \label{eq:master1}
\end{align}
\end{ceqn}
where $\hat{\rho}(t)$ and $\hat{H}$ are the spin density and Hamiltonian operators, respectively. [·,·] denotes the commutator. 

\begin{figure}[ht!]
 \includegraphics[width=0.8\linewidth]{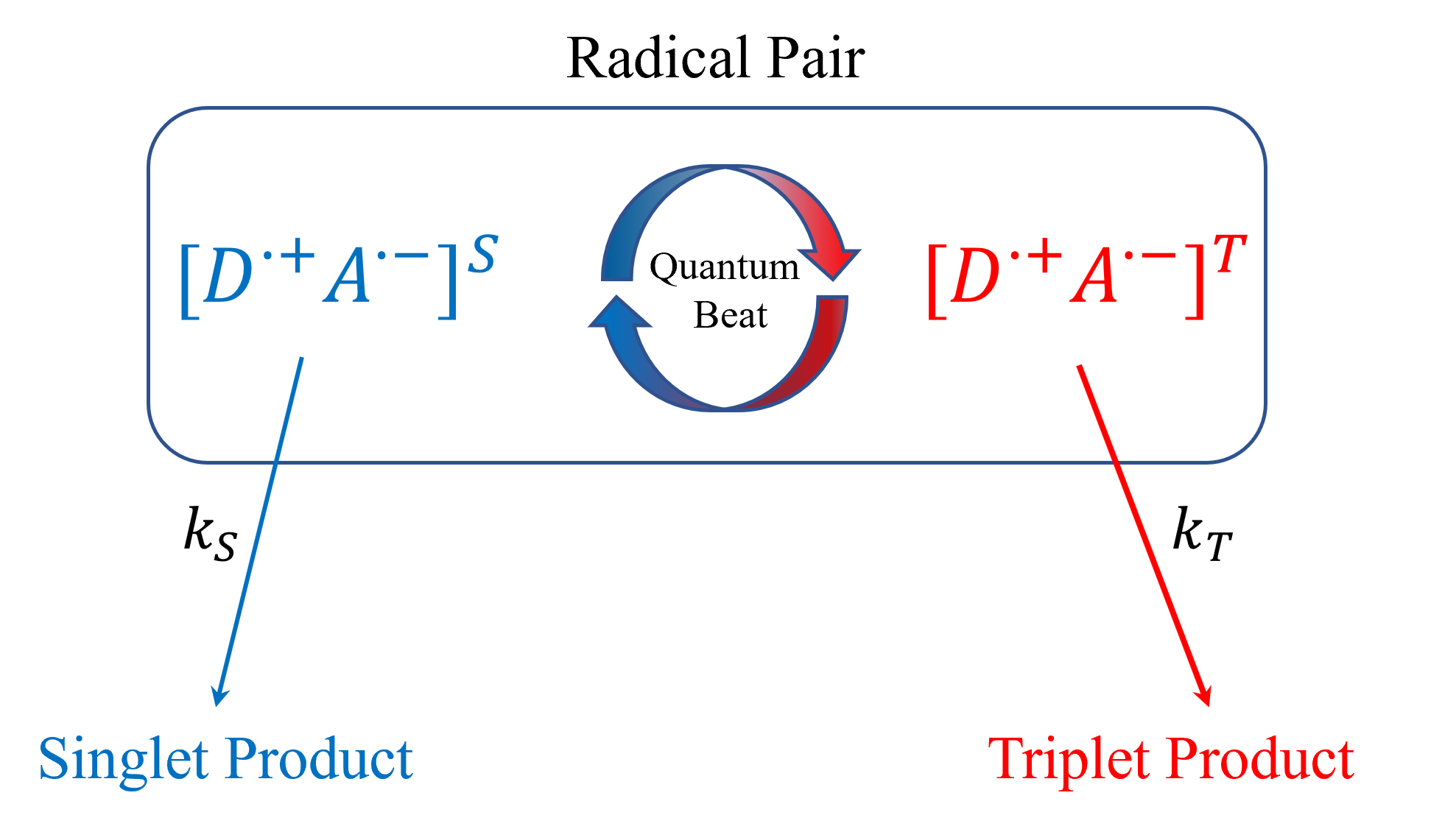}
 \caption{A simple schematic presentation of donor (D)-acceptor (A) radical pair reaction undergoing intersystem crossing between singlet (S) and triplet (T) states. Each state takes different chemical pathways via distinct reaction rates to produces S and T products with $k_S$ and $k_T$, respectively, for S and T states.} 
\label{fig:schem-quantum-beat}
\end{figure}

For instance, the probability of finding the radical pairs in singlet states at some later time is determined by Hamiltonian using Eq. \ref{eq:master1}:

\begin{ceqn}
\begin{equation}
 \expval{\hat{P}^S}(t) =\Trace [\hat{P}^S \hat{\rho}(t)], 
 \label{eq:yield}
\end{equation}
\end{ceqn}
where Tr is trace. 

The probability $\expval{\hat{P}^S}(t)$ depends on other contributions, including kinetic reactions, spin relaxation, vibration and rotation of radical pairs, which can be introduced to Eq. \ref{eq:master1}.


\subsubsection{Static magnetic field}

Static magnetic field effects have been extensively studied in the context of birds' magnetosensitivity \cite{Rodgers2009,Xu2021}. However, the applications of these models can extended to other magnetic field effects reviewed in Section \ref{sec:SMF}. Assuming that the spin of the radical pairs start off from a singlet state, Eq. \ref{eq:yield} can be rewritten as:
\begin{ceqn}
\begin{equation}
 \expval{\hat{P}^S}(t)=\frac{1}{M} \sum_{m}^{4M} \sum_{n}^{4M} \abs{\bra{m}\hat{P}^S\ket{n}}^2 \cos{([\omega_m-\omega_n]t)},
\end{equation}
\end{ceqn}
where $\ket{m}$ and $\ket{n}$ are eigenstates of $\hat{H}$ with corresponding eigenenergies of $\omega_m$ and $\omega_n$, respectively. 

Spin relaxation can be introduced phenomenologically \cite{Bagryansky2007,Hore2019} such that: 

\begin{ceqn}
\begin{equation}
 \expval{\hat{P}^S}(t) \longrightarrow \frac{1}{4} - \big(\frac{1}{4}-\expval{\hat{P}^S}(t)\big) e^{-rt},
\end{equation}
\end{ceqn}
where $r$ denotes the spin relaxation rate. Following the work of Timmel et al. \cite{Timmel1998}, the chemical fate of the radical pair can be modelled separating spin-selective reactions of the singlet and triplet pairs, as shown in Figure \ref{fig:schem-quantum-beat}. For simplicity it is assumed that $k=k_S=k_T$, where $k_S$ and $k_T$ are the singlet and triplet reaction rates, respectively. The final singlet yield, $\Phi_S$, for periods much greater than the radical pair lifetime reads as follows:

\begin{ceqn}
\begin{align}
 \Phi_S & = k \int_{0}^{\infty} \expval{\hat{P}^S}(t) e^{-kt} dt  ={}\frac{1}{4}-\frac{k}{4(k+r)}+\frac{1}{M} \sum_{m}^{4M} \sum_{n}^{4M} \abs{\bra{m}\hat{P}^S\ket{n}}^2 \frac{k(k+r)}{(k+r)^2+(\omega_m -\omega_n)^2},
  \label{eq:SY}
\end{align}
\end{ceqn}
where the fractional triplet yield can be calculated as $\Phi_T=1-\Phi_S$. \par

In Section \ref{sec:RPM-Brain}, we briefly review recent studies that suggest the radical pair mechanism may explain xenon-induced anesthesia, lithium effects on hyperactivity, magnetic field and lithium effects on circadian clock, and hypomagnetic field effects on neurogenesis and microtubule reorganization (See Section \ref{sec:RPM-Brain}). In these studies, for simplicity, only Zeeman and isotropic hyperfine interactions are considered. For a pair of radicals, the Hamiltonian reads: 

\begin{ceqn}
\begin{equation}
 \hat{H}=\omega \hat{S}_{A_{z}}+\mathbf{\hat{S}}_A.\sum_i^{N_A} a_{A_i} \mathbf{\hat{I}}_{A_i}+\omega \hat{S}_{D_{z}}+\mathbf{\hat{S}}_D.\sum_i^{N_D} a_{D_i} \mathbf{\hat{I}}_{D_i},
 \label{eq:ham}
\end{equation}
\end{ceqn}
where $\mathbf{\hat{S}}_A$ and $\mathbf{\hat{S}}_D$ are the spin operators of radical electrons on \ch{A^{.-}} and \ch{D^{.+}}, respectively, $\mathbf{\hat{I}}_A$ and $\mathbf{\hat{I}}_D$ are the nuclear spin operators on the acceptor and donor radical molecule, $a_{A}$ and $a_{B}$ are the isotropic hyperfine coupling constants, $N_A$ and $N_D$ are the number of nuclei coupled to electron $A$ and $D$, respectively, and $\omega$ is the Larmor precession frequency of the electrons due to the Zeeman effect. \par

\subsubsection{Hypomagnetic field}

Although hypomagnetic fields belong to the static magnetic field category, the effects due to extremely low magnetic field are often particularly significant compared to other magnetic field effects. 



\begin{figure}[ht!]
 \includegraphics[width=0.8\linewidth]{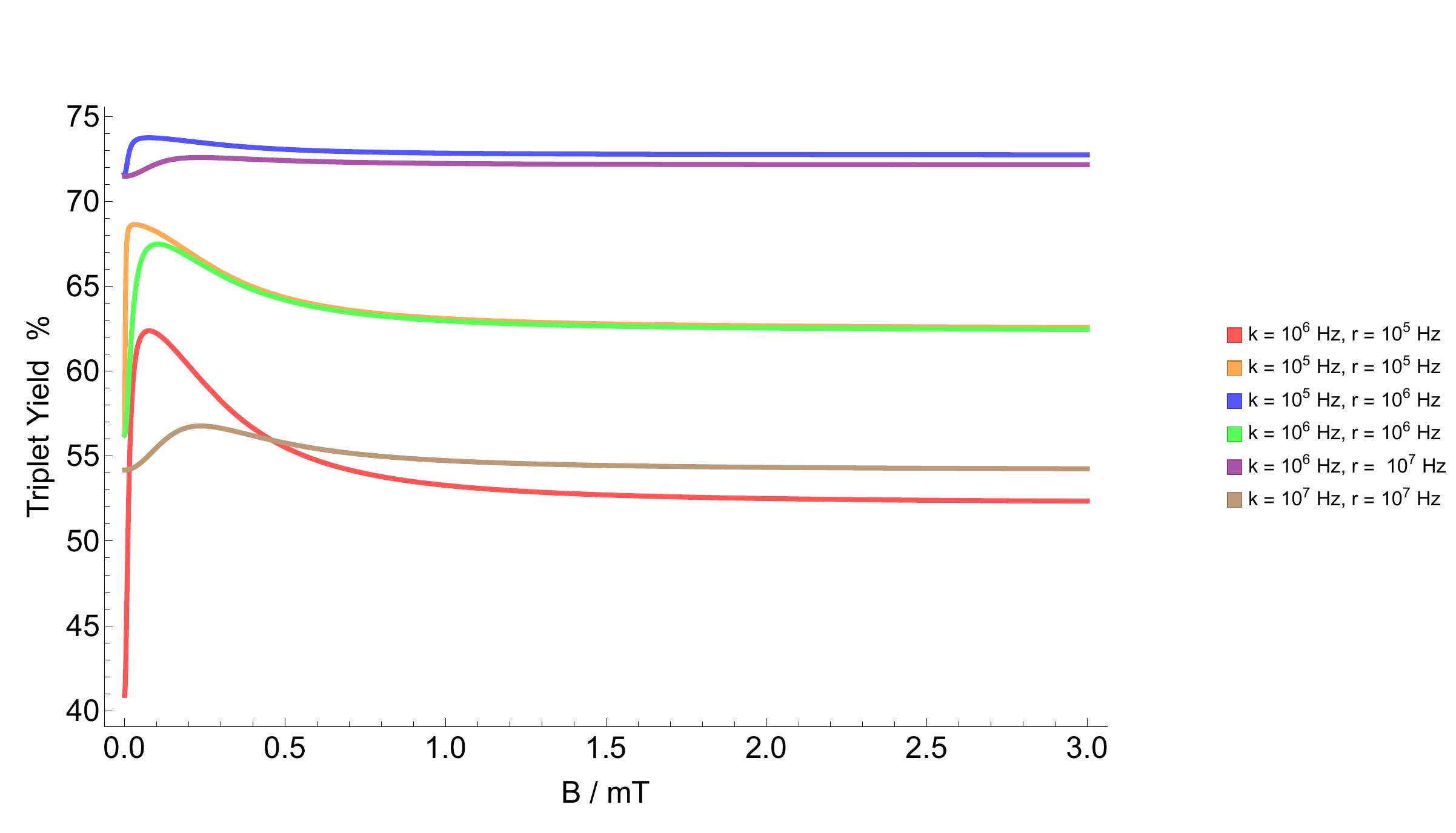}
 \caption{Triplet yield vs applied magnetic field for different reaction and spin relaxation rates for a simple model of a radical pair. In this model, one of the radicals is coupled with a nucleus with a hyperfine coupling constant of 0.3 mT. For different values of the rates, one can see a pronounced dip near zero field, together with a maximum close to the value of the geomagnetic field (around 0.05 mT)} 
\label{fig:HMF}
\end{figure}

Using Eq. \ref{eq:SY}, it can be shown that for different relaxation and reactions rates, the hypomagnetic field effects are significant, as show in Fig. \ref{fig:HMF}.

\subsubsection{Extremely low-frequency magnetic field}

Given the short lifetime of radical pairs compared to the low frequency of applied magnetic field, in general, the extremely low-frequency magnetic field can be treated as static during the lifetime of a radical pair \cite{Scaiano1994,Hore2019}. Depending on the phase of oscillation, $\alpha\in (0,\pi)$, each radical pair therefore experiences a different, effectively static, magnetic field whose field strength is $B$. The net effect of the oscillating field is an average over $\alpha$, such that:

\begin{ceqn}
\begin{equation}
B(t)=B_0+B_1(t) \Longrightarrow B(t)\equiv B=B_0+B_1 \cos{\alpha},
\label{eq:osc1}
\end{equation}
\end{ceqn}

\begin{ceqn}
\begin{equation}
    \overline{\Phi_S(B_0,B_1)}=\frac{1}{\pi}\int_0^\pi\Phi_S(B) d\alpha,
    \label{eq:osc2}
\end{equation}
\end{ceqn}
where $B_0$ and $B_1$ indicate the static magnetic field and the amplitude of the oscillating magnetic field, receptively. Such theoretical model can be applied to the magnetic field effects reviewed in Section \ref{sec:LFOMF}.\par

\subsubsection{Medium/high-frequency magnetic field}

For the cases of medium/high frequency magnetic fields, a general approach is to integrate Eq. \ref{eq:master1}, using, for example, a 4th order Runge-Kutta scheme. It is shown that high-frequency magnetic effects can be accounted for by the radical pair mechanism \cite{Canfield1995,Timmel1996,Hiscock2016Floquet}. For instance, if the magnetic field has the following form:

\begin{ceqn}
\begin{equation}
B(t)=B_0 \hat{k}+B_1 [\cos{\omega t} \hat{i} +\sin{\omega t} \hat{j} ],
\label{eq:osc3}
\end{equation}
\end{ceqn}
the corresponding Hamiltonian can be transformed into a rotating reference frame where it becomes a time-independent Hamiltonian \cite{Canfield1994}. To do so, one could use a unitary transformation matrix:
\begin{ceqn}
\begin{equation}
T(t)=e^{i(\hat{S}_{A_{z}}+\hat{I}_{A_{z}}+\hat{S}_{B_{z}}+\hat{I}_{B_{z}})\omega t},
\end{equation}
\end{ceqn}
such that:
\begin{ceqn}
\begin{equation}
    H^{'} =i\hbar \Dot{T}(t)T^{-1}(t)+T(t)H(t)T^{-1}(t),
\end{equation}
\end{ceqn}

Where $H^{'}$ is the time-independent Hamiltonian and $\Dot{T}(t)$ is the time derivative of $T(t)$. After some algebra, one can obtain: 
\begin{ceqn}
\begin{equation}
    H^{'} =g \mu_B \sum^{2}_{j=1}(B_0 S_{jz}+B_1 S_{jx}+a_j \mathbf{\hat{S}}_j.\mathbf{\hat{I}}_j)-\omega\sum^{2}_{j=1} (S_{jz}+I_{jz}).
\end{equation}
\end{ceqn}

\subsection{Candidate radical pairs}\label{sec:RP-Candidates}

It is now well known that in biology electron-transfer reactions can take place at reasonable rates even when the reactants are separated far beyond “collisional” distances \cite{McLendon1992,Moser1995}. A radical pair can be formed by breaking a chemical bond or electron-transfer between two molecules. Electron transfer between proteins is facilitated by formation of a complex of the reacting proteins, which may be accompanied by conformational changes in the proteins. For that, the reactants must reach each other to build up the coupling of their electronic orbitals. The most used approach to rationalize and predict the rate of electron transfer processes is Marcus Electron Transfer theory \cite{Marcus1985}. Determining realistic radical pair candidates for the magnetosensitivity of physiological function, however, is still an interesting challenge. Here, we briefly review a few plausible radical pairs that maybe be relevant for the magnetosensitivity in biology. 

\subsubsection{Cryptochrome-based radical pairs}

In the context of songbird avian magnetoreception, the cryptochrome proteins are the canonical magnetosensitive agent \cite{Dodson2013,Mouritsen2012,Xu2021}. It is thought that, in cryptochromes and photolyases, photoreduction of FAD is through three consecutive electron transfers along a conserved triad of tryptophan (Trp) residues to give \ch{FAD^{.-}} and \ch{TrpH^{.+}} approximately 2 nm distant from each other \cite{Cailliez2016,Giovani2003,Mller2015,Zeugner2005}. In cryptochrome-4a, sequentially four radical pair states are formed by the progressive transfer of an electron along a chain of four tryptophan residues to the photo-excited flavin. In a recent study, Hore and co-workers suggest that, based on spin dynamics, while the third radical pair is mainly responsible for magnetic sensing, the fourth could enhance initiation of magnetic signalling particularly if the terminal tryptophan radical can be reduced by a nearby tyrosine (Tyr) \cite{Wong2021four}. They concluded that this arrangement may play an essential role in sensing and signalling functions of the protein. It is also suggested that tyrosine can be the donor instead of the fourth Trp \cite{Giovani2003}. It is also found based on spin dynamics analysis that a radical pair in the form form of [\ch{FAD^{.-}} and \ch{Tyr^{.}}] can provide sensitivity to the direction of the magnetic field \cite{Hong2020}.

There is also a consensus that there exist alternative radical pairs beyond [\ch{FAD^{.-}}...\ch{TrpH^{.+}}]. In 2009, Ritz and Schulten showed that exposure to low-intensity oscillating magnetic fields disoriented European robins \cite{Ritz2000}. Interestingly the frequency of the applied magnetic field in that experiment was equal to the Larmor frequency ($\sim$ 1.4 MHz) of a free electron spin in the geomagnetic field. Magnetic fields with the same amplitude but different frequencies had much less impacts on the birds' magnetic compass. Theoretical analysis suggests that such phenomenon may be explained if one of the radicals were free from internal magnetic interactions \cite{Mller2011,Niener2014,Niener2013,Ritz2009}, which implies that such an observation is not compatible with the radical pair model based on [\ch{FAD^{.-}}...\ch{TrpH^{.+}}]. A lot of evidence suggests that the superoxide radical is the most plausible radical under such circumstances \cite{Ritz2000,Maeda2008,Hogben2009,Solovyov2009,Mller2011,Lee2014,vanWilderen2015}. This is also consistent with animal magnetoreception in dark \cite{Netuil2021,Hiscock2019,Player2019}, as it was suggested that during the backreaction, a radical pair is formed between flavin and an \ch{O2} and that the radical pair reaction responds significantly to reorientation in the geomagnetic field \cite{Solovyov2008,Maeda2008,Ritz2009,Bouly2007,Prabhakar2002}. Such a radical pair could be generated without further absorption of light in the form of [\ch{FADH^{.}}...\ch{O2^{.-}}]. However, deciding the more realistic radical pair between [\ch{FADH^{.}}...\ch{O2^{.-}}] and [\ch{FAD^{.-}}...\ch{TrpH^{.+}}] to explain avian magnetoreception is still a matter of active debate \cite{Player2019,Schwarze2016,Engels2014,Hiscock2017}.

\subsubsection{Beyond cryptochrome-based radical pairs}

\begin{figure}
     \begin{subfigure}{0.55\linewidth}
        \includegraphics[width=1\textwidth]{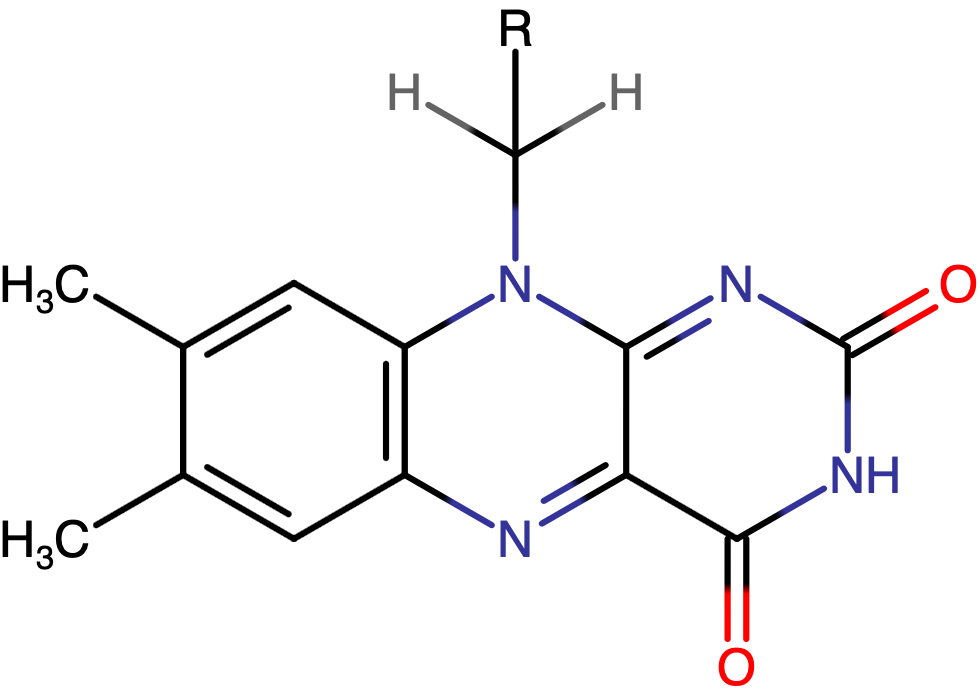}
        \caption{}
    \end{subfigure}
    \hfill
\begin{minipage}{0.7\linewidth}
    \begin{subfigure}{\linewidth}
\includegraphics[width=1\textwidth]{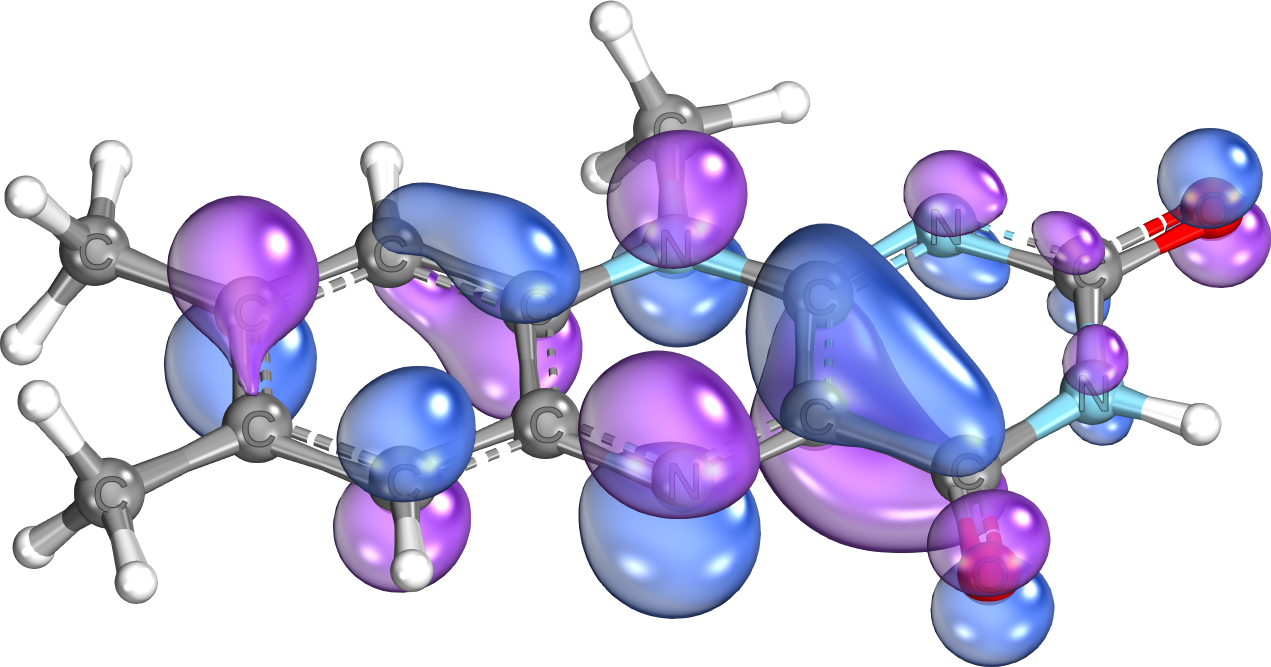}
        \caption{}
    \end{subfigure}
\end{minipage}%

\caption{Molecular structure and orbitals of the flavin radical. (a) Structure of flavin adenine dinucleotide (FAD). R denotes the adenosine diphosphate group and the rest of the ribityl chain. (b) Representations of the molecular orbitals that contain the unpaired electron in a flavin anion radical. Blue and purple indicate parts of the wave function with opposite signs. ORCA packaged used to calculate the HOMO using PBE0/def2-TZVP \cite{Neese2011}. Image rendered using \href{http://www.iboview.org/}{IboView [v20211019-RevA]}.}  
\label{fig:FAD}
\end{figure}

Flavin-dependent enzymes are ubiquitous in biology. The isoalloxazine ring of the flavin cofactor can undergo thermally driven redox chemistry, as shown in Figure \ref{fig:FAD}. The different redox states of flavin play essential roles in various electron transfer processes and consequently are crucial for a variety of important biological functions, including energy production, oxidation, DNA repair, RNA methylation, apoptosis, protein folding, cytoskeleton dynamics, detoxification, neural development, biosynthesis, the circadian clock, photosynthesis, light emission and biodegradation \cite{Moser1995,Massey2000,Joosten2007,Romero2018,Walsh2013,Fraaije2000,Vanoni2013,Vitali2016,Hamdane2016,Udhayabanu2017,Zwang2018,Husen2019,Lukacs2022}. Different forms of transient radical pair intermediates can be created during reactions catalysed by flavin-dependent enzymes, including [\ch{FADH^{.}}...\ch{O2^{.-}}] \cite{Gbicki2004,Fukuzumi2008,Yuasa2008}.

Although cryptochrome is the main protein for the avian magnetoreception, there exist many observation challenges for the canonical cryptochrome-centric radical pair mechanism. In a recent work, Bradlaugh and co-workers observed that the FAD binding domain and the Trp chain in cryptochrome are not required for magnetic field responses at single neuron and organismal level. They further reported that increase in FAD intracellular concentration enhanced neuronal sensitivity to blue light in the presence of a magnetic field. The authors concluded that the magnetosensitivity in cells may be well explained based on non-cryptochrome-dependent radical pair models \cite{Bradlaugh2021}.

It is known that near the tetrodotoxin binding site in \ch{Na^{+}} channels there are tryptophan residues. Similarly, in the pore-forming region of voltage-sensitive \ch{Na^{+}} channels, tyrosine and tryptophan residues are located. It is suggested that gating these channel proteins may depend on the electron transfer between these residues, and hence formation of radicals \cite{Lee1992}. This form of electron transfer is also proposed to play a key role in DNA photolyase \cite{Aubert1999}.

Many physiological and pathological processes involve protein oxidation \cite{Stadtman2006}, icluding important residues such as Trp, Tyr, histidine (His), and proline (Pro). It is known that a radical pair in the form of [\ch{TyrO^{.}}...\ch{O2^{.-}}] can be created \cite{HoueLvin2015}. The superoxide radical can also be formed in a spin correlated manner with other partners, including tetrahydrobiopterin \cite{Eberlein1984,Adams1997,Roberts2013}. In addition, it was shown that an electron transfer process can occur between Trp and superoxide \cite{McCormick1978,Saito1981}. It was also suggested that in phosphoglycerate kinase phosphorylation a radical pair [\ch{RO^{.}}...\ch{Mg}(\ch{{H2O})n^{.+}}] complex can be formed \cite{Buchachenko2005}.

\section{Studies of the potential role of radical pairs in the brain} \label{sec:RPM-Brain}
In this section, we briefly review recent studies that suggest that the radical pair mechanism may explain isotope effects in xenon-induced anesthesia, and lithium effects on hyperactivity, magnetic field and lithium effects on the circadian clock, and hypomagnetic field effects on neurogenesis and microtubule reorganization.

\subsection{Xenon anesthesia}\label{sec:RPM-Xe}

Xenon is a well-known general anesthetic used for several species, including \textit{Drosophila}, mice, and humans \cite{Franks1998}. Despite its simple structure (a single atom), the exact underlying mechanism by which it exerts its anesthetic effects remains unclear. Turin et al. showed that when xenon acts anesthetically on \textit{Drosophila}, specific electron spin resonance (ESR) signals can be observed \cite{Turin2014}. The same authors proposed that the anesthetic action of xenon may involve some form of electron transfer. Moreover, Li et al. showed experimentally that isotopes of xenon with non-zero nuclear spin had reduced anesthetic potency in mice compared with isotopes with no nuclear spin \cite{Li2018}. These findings are consistent with the hypothesis of radical pair creation in xenon-induced anesthesia. \par

\begin{figure}
        \begin{subfigure}{0.4\linewidth}
            \includegraphics[width=1\textwidth]{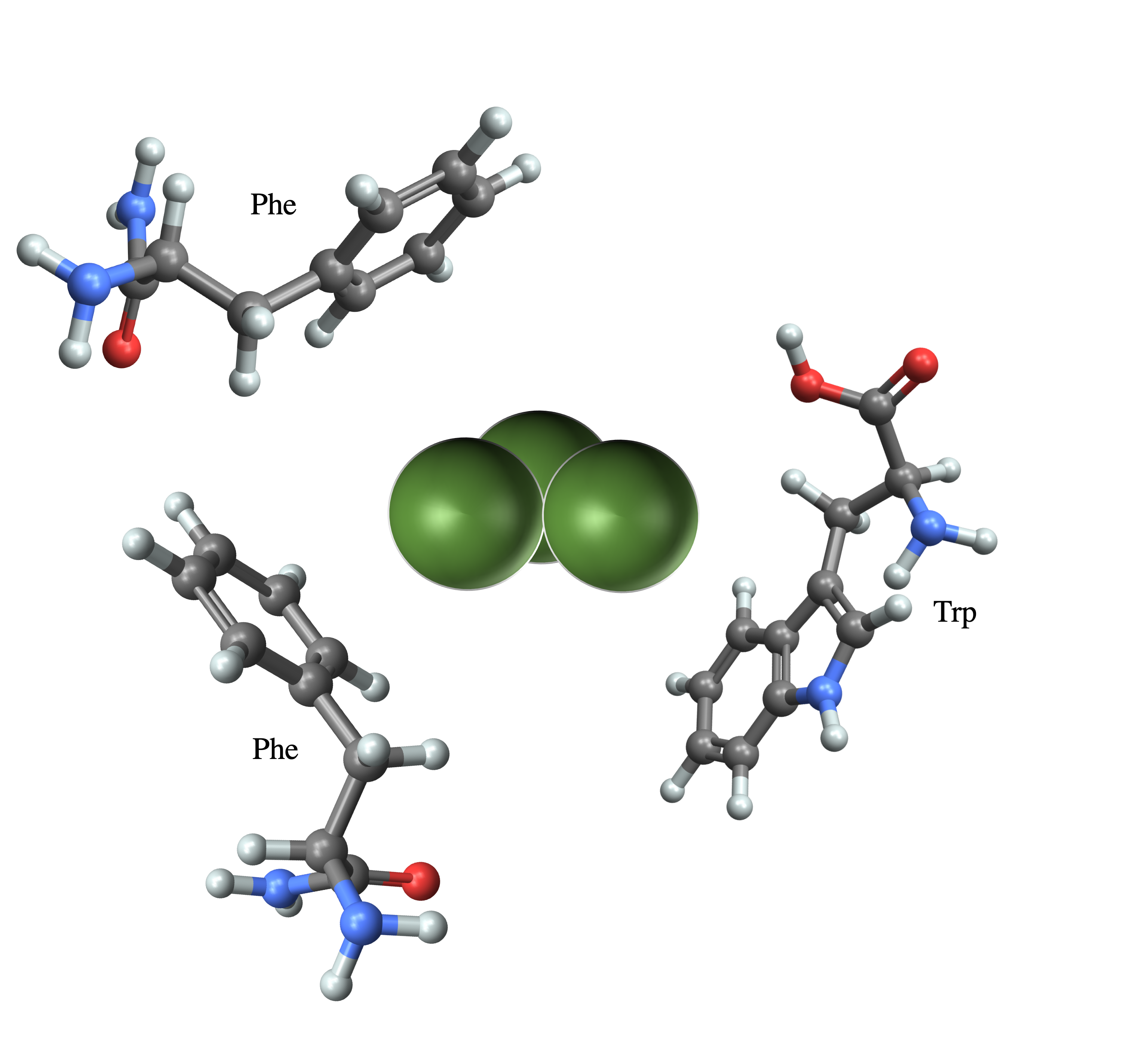}
            \caption{}
            \label{fig:NMDA-Xe-Schem}
     \end{subfigure}
     \hfill
\begin{minipage}{0.54\linewidth}
     \begin{subfigure}{\linewidth}
\includegraphics[width=1\textwidth]{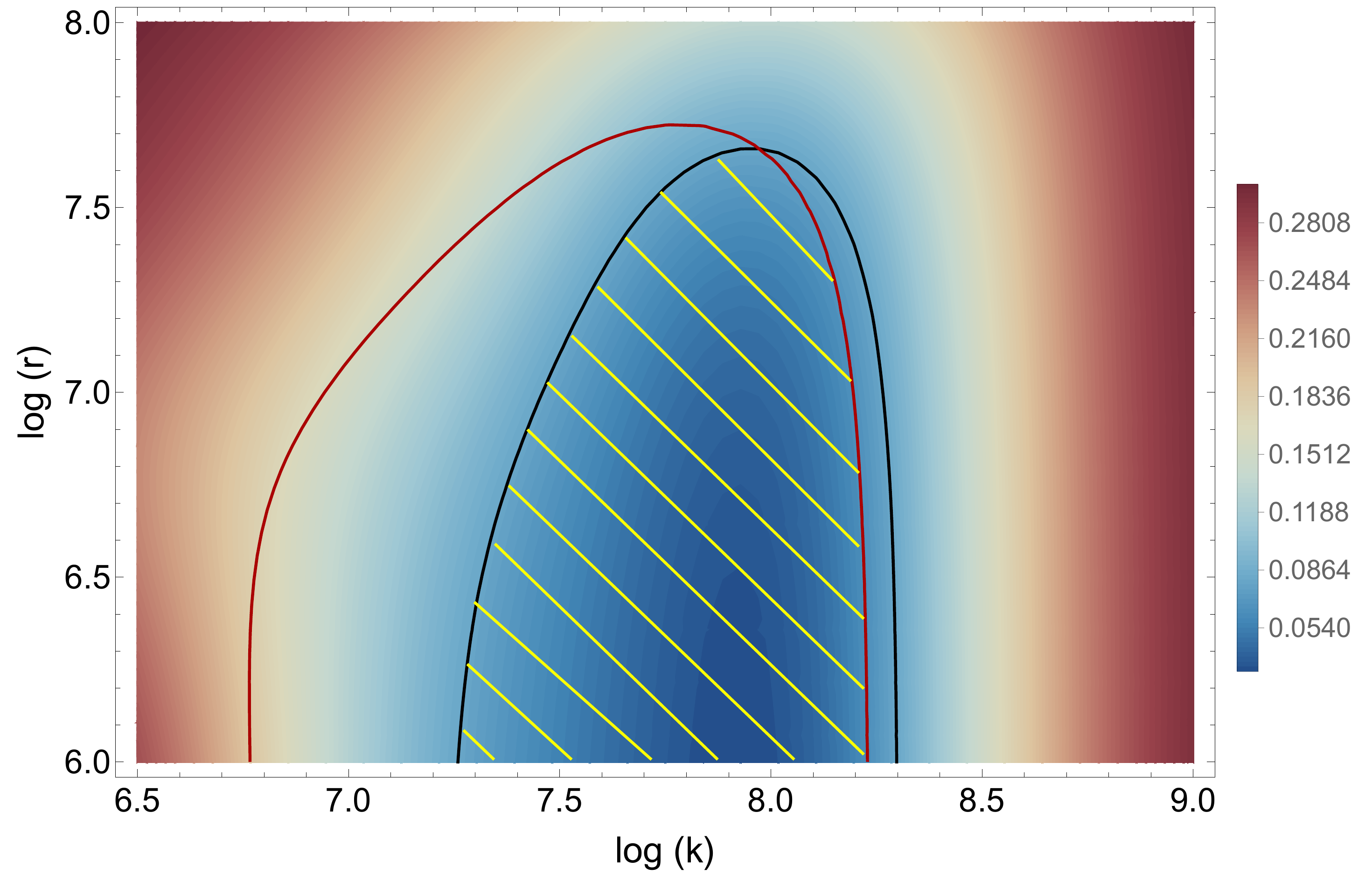}
            \caption{}
            \label{fig:Xe-Cntr}
     \end{subfigure}
\end{minipage}%
     \hfill
        \begin{subfigure}{0.5\linewidth}
             \centering
            \includegraphics[width=1\textwidth]{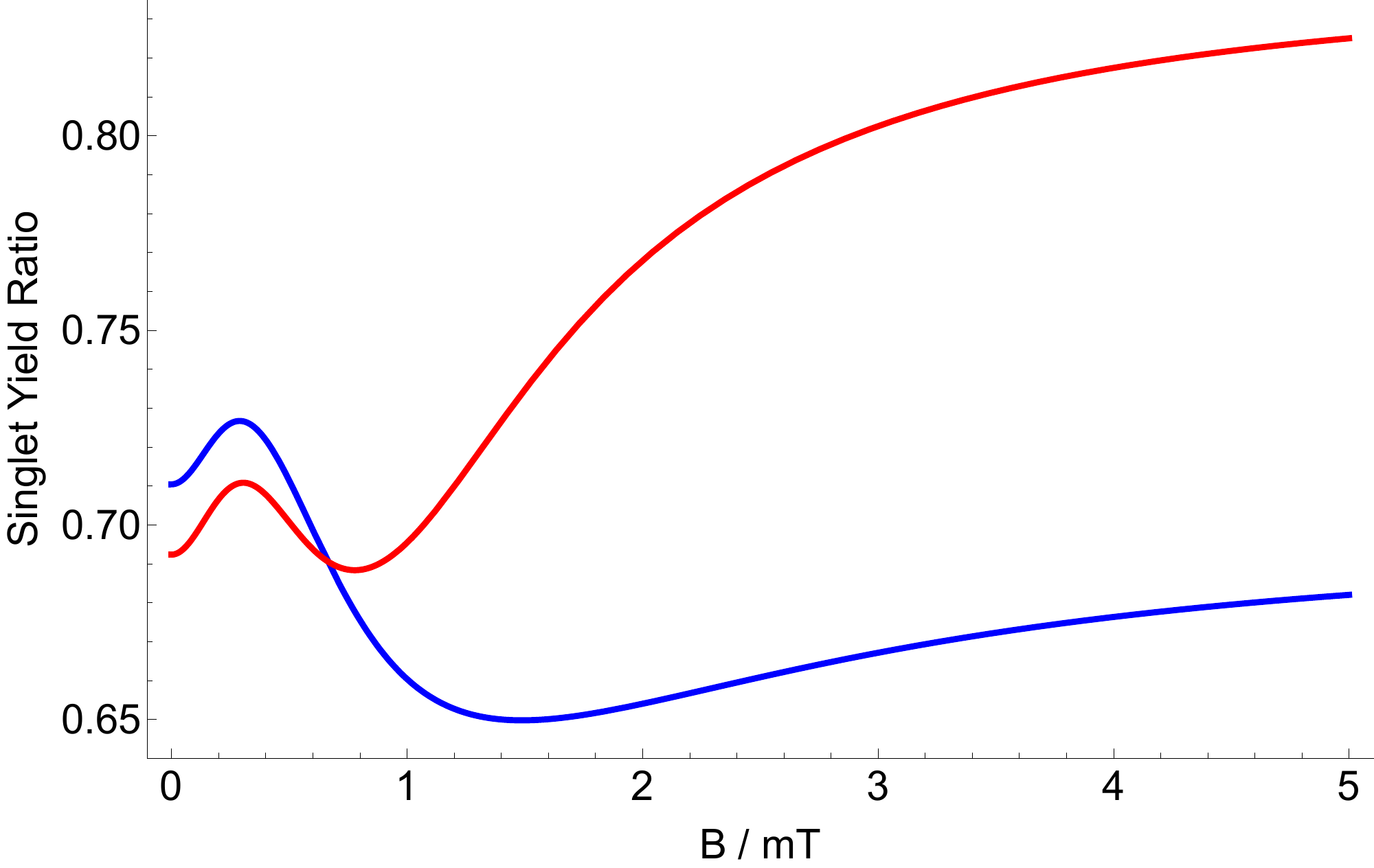}
            \caption{}
            \label{fig:Xe-SYR}
     \end{subfigure}
 
\caption{Radical pair explanation for isotope effects in xenon-induced anesthesia. (a) Schematic presentation of the interaction of xenon (Green spheres) with aromatic rings of tryptophan (Trp) and phenylalanine (Phe) at the glycine binding site of the NMDAR \cite{Armstrong2012}. (b) The dependence of the agreement between relative anesthetic potency and singlet yield ratio on the relationship between relaxation rate, $r$, and reaction rate, $k$. The radical pair model can explain the experimentally derived relative anesthetic potency of xenon, shaded in yellow. (c)  Predicted dependence of the anesthetic potency as given by the singlet yield ratio, based on the radical pair model of \ch{^{129}Xe}/\ch{^{130}Xe} (Blue) and \ch{^{131}Xe}/\ch{^{130}Xe} (Red) on an external magnetic field \cite{Smith2021}.}
\label{fig:Xe}
\end{figure}

Franks and co-workers identified NMDA subtype of glutamate receptor \cite{Traynelis2010} as a target for xenon anesthesia\cite{Franks1998,deSousa2000}. They further showed that xenon exerted its effects by inhibiting NMDA receptors (NMDARs) by competing with the coagonist glycine at the glycine-binding site on the GluN1 subunit \cite{Dickinson2007}. Subsequently, the same group identified that xenon interacts with a small number of amino acids at the predicted binding site of the NMDAR \cite{Armstrong2012}. Using Grand Canonical Monte Carlo method, they showed that xenon at the binding site can interact with tryptophan and phenylalanine, as shown in Fig.\ref{fig:NMDA-Xe-Schem}. However, due to redox inactivity, it is highly unlikely that phenylalanine can be involved in the electron transfer process \cite{Byrdin2007,Li1991}. Meanwhile, tryptophan is redox active and hence can feasibly be involved in electron transfer and hence the formation of radical pairs, as seen in the context of cryptochrome magnetoreception \cite{Hore2016}. In addition, it is known that tryptophan residues of the NMDAR play key roles in channel gating \cite{Williams933,Buck2000}. Moreover, exposure to low-intensity magnetic fields activates the NMDAR \cite{Manikonda2007,Salunke2013,zgn2019}.

It is also known that reactive oxygen species are implicated in the activation of the NMDARs \cite{Furukawa2003,Dickinson2007,Aizenman1990,Girouard2009,Betzen2009,Dukoff2014}. Moreover, Turin and Skoulakis \cite{Turin2018} reported that oxygen gas was necessary for observing spin changes during xenon-induced anesthesia in \textit{Drosophila}. Motivated by these observations, the authors \cite{Smith2021} suggested that the electron transfer related to xenon’s anesthetic action that is evidenced by Turin et al. \cite{Turin2014} plays a role in the recombination dynamics of a naturally occurring [\ch{Trp^{.+}}...\ch{O2^{.-}}] radical pair (See Section \ref{sec:RP-Candidates} for further discussion). Using Eqs. \ref{eq:SY} and \ref{eq:ham}, they showed that for isotopes of xenon with a non-zero nuclear spin, this nuclear spin couples with (at least one of) the electron spins of such a radical pair, affecting the reaction yields of the radical pair and hence xenon’s anesthetic action. The radical pair was assumed to start off from a singlet state. Such a mechanism is consistent with the experimental results of Li et al. \cite{Li2018} that xenon isotopes with non-zero nuclear spin have reduced anesthetic potency compared to isotopes with zero nuclear spin. The authors also provide an experimental test for the validity of their model. It predicts that under a static magnetic field the anesthetic potency of xenon may be significantly different than that observed by Li et al. \cite{Li2018}, as shown in Fig.\ref{fig:Xe-SYR}.

\subsection{Lithium effects on hyperactivity}\label{sec:RPM-Li}

Lithium (Li) is the most well known treatment for bipolar illness \cite{Berridge1989,Klein1996,Geddes2013,Grande2016,Vieta2018,Burdick2020}. Despite its frequent clinical use, the mechanism by which Li exerts its effects remains elusive \cite{Jope1999}. Ettenberg and co-workers \cite{Ettenberg2020} showed that Li effects on the manic phase in rats are isotope-dependent. They used sub-anesthetic doses of ketamine to induce hyperactivity which was then treated with lithium. They observed that \ch{^6Li} produced a longer suppression of mania compared to \ch{^7Li}. Further, there is considerable amount of evidence that oxidative stress \cite{Sies2020} is implicated in both bipolar disorder \cite{Berk2011,Andreazza2008,Yumru2009,Steckert2010,Wang2009,Salim2014,Ng2008,Lee2013,Brown2014} and its Li treatment \cite{MachadoVieira2011,MachadoVieira2007,frey2006effects,deSousa2014}. \par

\begin{figure}
        \begin{subfigure}{0.5\linewidth}
            \includegraphics[width=1\textwidth]{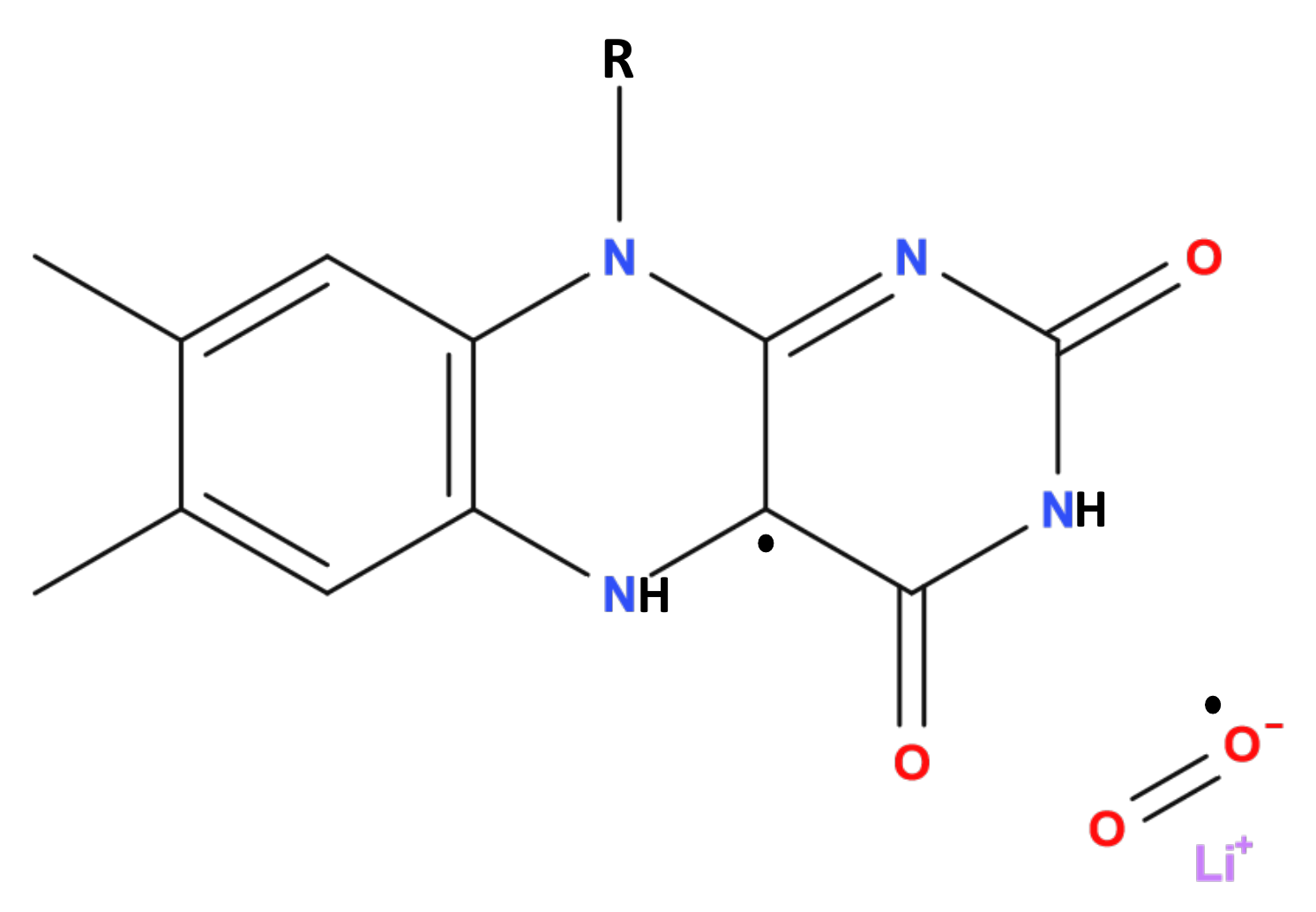}
            \caption{}
            \label{fig:Li-Schem}
     \end{subfigure}
     \hfill
\begin{minipage}{0.53\linewidth}
     \begin{subfigure}{\linewidth}
\includegraphics[width=1\textwidth]{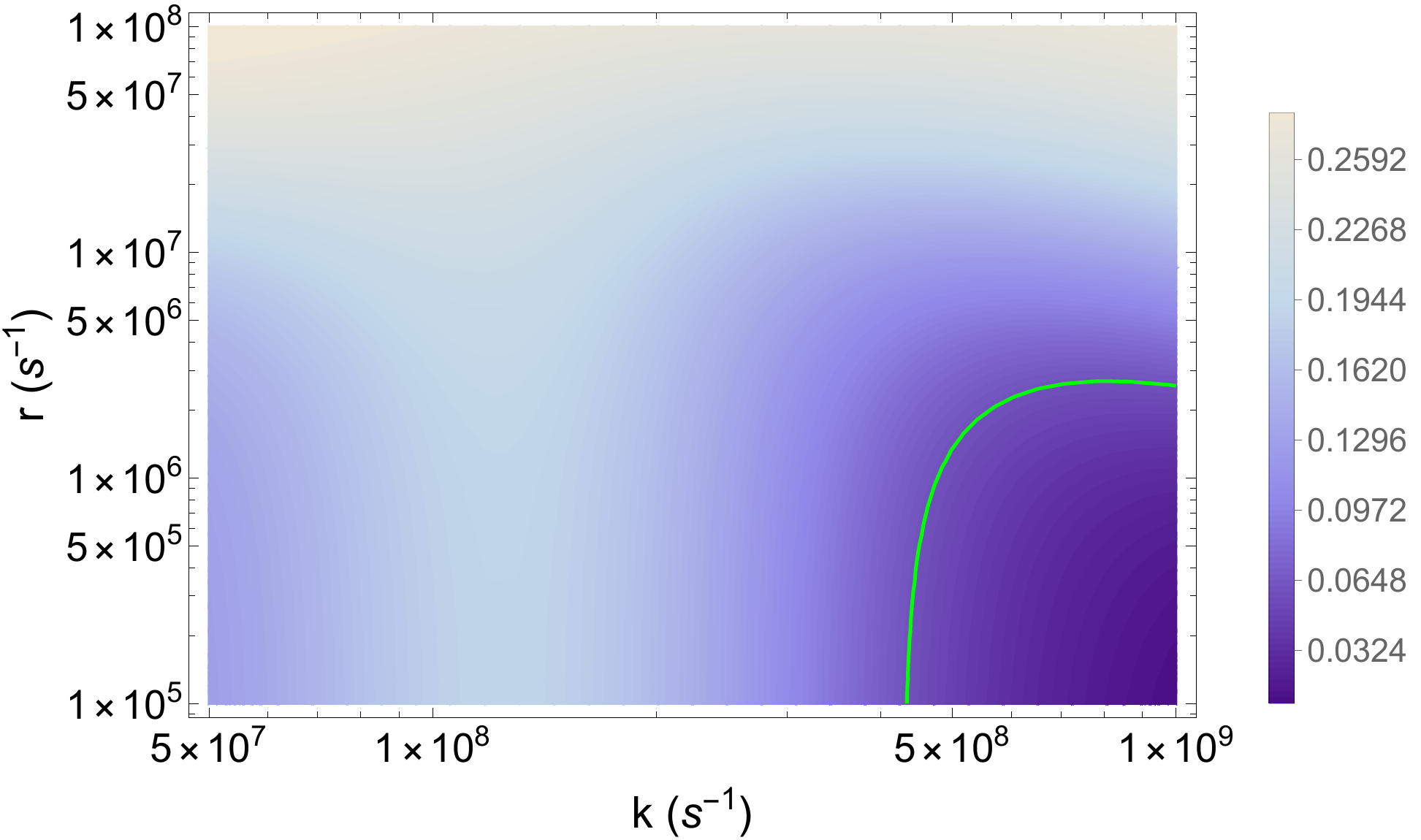}
            \caption{}
            \label{fig:Li-Cntr}
     \end{subfigure}
\end{minipage}%
     \hfill
        \begin{subfigure}{0.53\linewidth}
             \centering
            \includegraphics[width=1\textwidth]{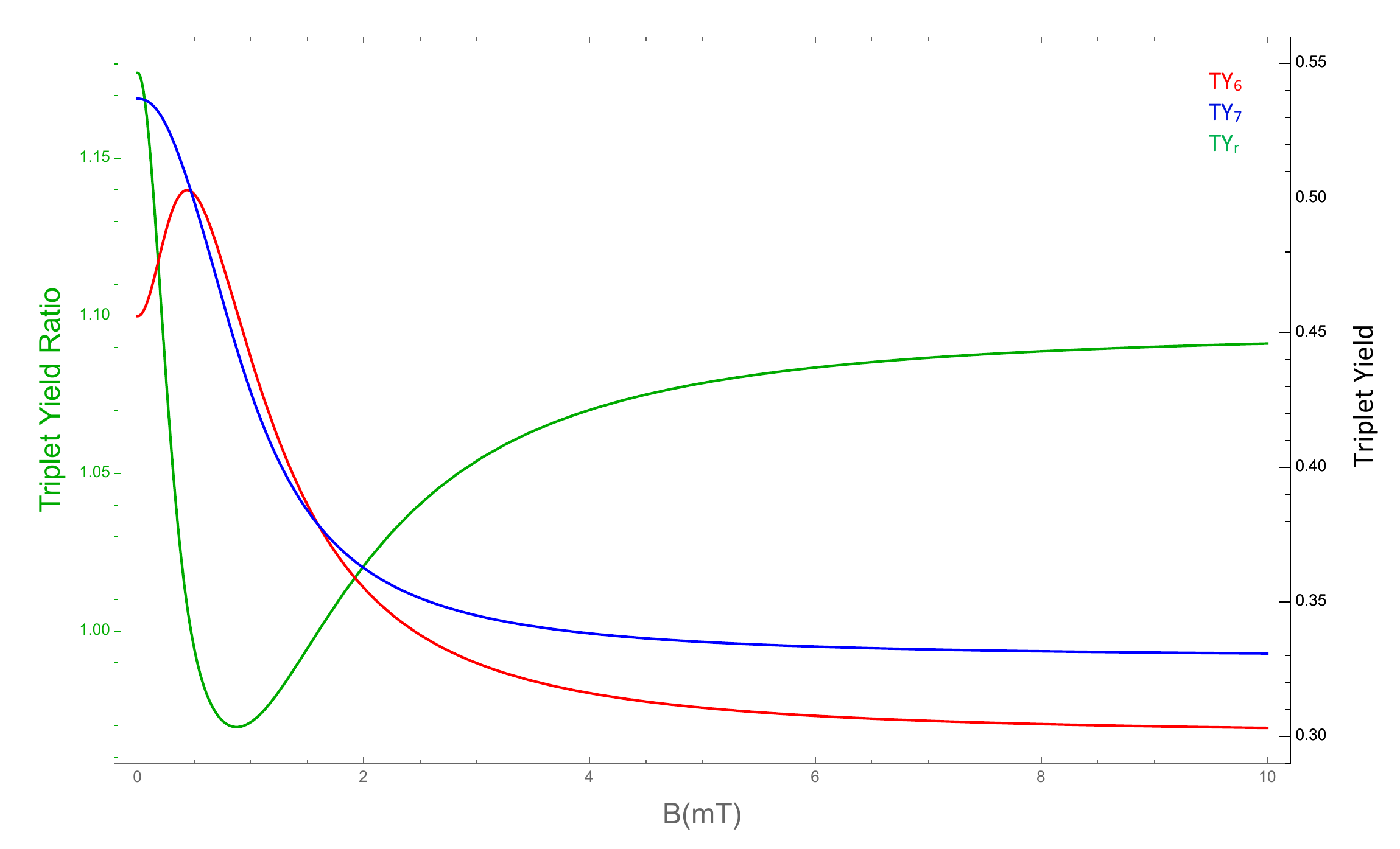}
            \caption{}
            \label{fig:Li-TYR}
     \end{subfigure}
 
\caption{Radical pair explanation for isotope effects in lithium treatment for hyperactivity. (a) Flavinsemiquinone (\ch{FADH^{.}}) and lithium superoxide radical pair (\ch{Li^{+}}...\ch{O2^{.-}}). (b) The dependence of the agreement between the total traveled distance ratio, $TD_r$, and the triple yield ratio, $TY_r$ of \ch{^7 Li} over \ch{^6 Li} on the radical pair reaction rate, $k$, and the radical pair spin-coherence relaxation rate, $r$. The green line indicates the ranges smaller than the experimental uncertainty. (c) The dependence of the triplet yield (Red, \ch{^6Li}; Blue, \ch{^7Li}) and triplet yield ratio \ch{^7Li}/\ch{^6Li} (Green)
on an external magnetic field, calculated based on the radical pair model \cite{Zadeh2021Li}.}
\label{fig:Li}
\end{figure}

Bipolar disorder is also correlated with irregularities in circadian rhythms \cite{Takahashi2008,Lee2019,Porcu2019,Fang2021}. In addition, it is well known that Li influences the circadian rhythms that are disrupted in patients with bipolar disorders \cite{Engelmann1973,Delius1984,Possidente1986,Klemfuss1992,Moreira2016,Geoffroy2017,McCarthy2018,Wei2018,Papiol2019,Sawai2019,Andrabi2019,Sanghani2020,Xu2021Li,Federoff2021}. Further, Osland et al. reported that Li significantly enhanced the expression of \textit{Per2} and \textit{Cry1}, while \textit{Per3}, \textit{Cry2}, \textit{Bmal1}, \textit{E4BP4} and \textit{Rev-Erb-}$\alpha$ expression was decreased \cite{Osland2010}. However, the exact mechanisms and pathways behind this therapy are incompletely known. It has been shown that Li can exert it effects via a direct action on the suprachiasmatic nucleus (SCN), a circadian pacemaker in the brain \cite{LeSauter1993,Abe2000,Yoshikawa2016,Vadnie2017}. Cryptochromes are key proteins for the circadian clock \cite{Horst1999} and SCN’s intercellular networks development, which subserves coherent rhythm expression \cite{Welsh2010}. Furthermore, it is also shown that cryptochrome is associated with bipolar disorder disease \cite{Lavebratt2010,Schnell2015,Hhne2020,Sokolowska2020}. In the context of animal magnetoreception, cryptochromes are the canonical magnetic sensing proteins (See Section \ref{sec:RP-Candidates}) \cite{Hore2016}, with flavin radicals playing a key role. Moreover, it has been shown that circadian rhythms are susceptible to magnetic fields at low intensities \cite{Brown1960,wever1970effects,BLISS1976,Close2014,Close2014b,Manzella2015,Bradlaugh2021,CONTALBRIGO2009,Bartos2019}, where cryptochromes \cite{yoshii2009cryptochrome,Fedele2014} are implicated. It has also been observed that cryptochromes play key roles in alteration of ROS levels through exposure to magnetic fields \cite{Sherrard2018,sherrard2018low,Pooam2018,Pooam2020}. Based on these facts, a new study suggests \cite{Zadeh2022CC} that Li’s nuclear spin influences the recombination dynamics of S-T interconversion in the naturally occurring [\ch{FADH^{.}}..\ch{O2^{.-}}] radical pairs (Fig.\ref{fig:Li-Schem}). These pairs are initially in singlet states, and due to the different nuclear spins, each isotope of Li alters these dynamics differently. Using Eqs. \ref{eq:SY} and \ref{eq:ham}, the authors showed that a radical pair model could provide results consistent with the experimental finding of Ettenberg and colleagues \cite{Ettenberg2020}. In that work, it was assumed that the fractional triplet yield of the radical pairs is correlated with lithium potency. They further predict a magnetic-field dependence of the effectiveness of lithium, which provides one potential experimental test of their hypothesis, as shown in Fig.\ref{fig:Li-TYR}. \par

Furthermore, the authors suggested that the proposed mechanism for Li effects is also plausible via different pathways. Li may exert it effect via competing with magnesium in inhibiting glycogen synthase kinase-3 (GSK-3) \cite{Stambolic1996,Ryves2001,Iwahana2004}, which is regulated by phosphorylation of inhibitory serine residues \cite{Fang2000,Jope2004,Beurel2015}. GSK-3 phosphorylates the clock components including PER2, CRY1, CLOCK, BMAL1, and REV-ERB$\alpha$ \cite{Yin2006,Iitaka2005,Harada2005,Yin2010,Sahar2010,Spengler2009,Besing2017,Breit2018}. In such cases, the radical pairs could be formed in a [\ch{RO^{.}}...\ch{Li}(\ch{{H2O})n^{.}}] complex (See Section \ref{sec:RP-Candidates}), where \ch{RO^{.}} is the protein oxy-anion, similar to Refs. \cite{Buchachenko2005,Buchachenko2009,Buchachenko2010,Buchachenko2014}.

\subsection{Magnetic field and lithium effects on circadian clock}\label{sec:RPM-CC}

The circadian clock is essential for the regulation of a variety of physiological and behavioral processes in nearly all organisms, including \textit{Neurospora} \cite{Loros1989}, \textit{Arabidopsis} \cite{Harmer2000}, \textit{Drosophila} \cite{Beaver2002}, mouse \cite{Peek2013}, and humans \cite{Ashbrook2019,Roenneberg2016,Takahashi2016}. It is known that the disruption of the circadian clock can be detrimental for many physiological functions, including depression \cite{Roybal2007,Taillard2021}, metabolic and cardiovascular diseases \cite{Crnko2019}, and cancer \cite{Battaglin2021,Sancar2021}. It is also known that the circadian clock controls physiological processes such as brain metabolism, reactive oxygen species homeostasis, hormone secretion, autophagy and stem cell proliferation, which are correlated to ageing, memory formation, neurodegenerative and sleep disorders \cite{Kondratova2012,AKondratova2010,Kyriacou2010,Liang2020,Maiese2021}. In \textit{Drosophila}, the circadian clock regulates the timing of eclosion, courtship, rest, activity, and feeding; it also influences daytime color \cite{Lazopulo2019} and temperature preference \cite{Allada2010}. Regardless of the differences in the molecular components of the circadian clocks, their organization, features, and the molecular mechanism that give rise to rhythmicity are very alike across organisms \cite{Tataroglu2014}.\par

Environmental zeitgebers such as light, food, and temperature can influence the circadian clock's rhythmicity\cite{Patke2019}. The circadian clock is also susceptible to magnetic field exposures \cite{BLISS1976,CONTALBRIGO2009,Marley2014,Fedele2014,Lewczuk2014,Vanderstraeten2015,Xue2021,Bartos2019,Manzella2015} (See also Sections \ref{sec:SMF-CC}). Yoshii et al. reported the effects of static magnetic fields with different intensities, [0, 150, 300, 500] $\mu$T, on the period changes of \textit{Drosophila}'s circadian clock under blue light illumination \cite{Yoshii2009}. They showed that the period was altered significantly depending on the strength of the magnetic field, with a maximum change at 300 $\mu$T. In that work, the geomagnetic field was shielded, and arrhythmic flies were excluded from the analysis. As discussed in Section \ref{sec:RPM-Li}, the disruption of the circadian clock is associated with bipolar disorders, for which Li is the first line treatment. Li's effects on bipolar disorder are isotope-dependent. Dokucu et al. \cite{Dokucu2005} reported that Li lengthened the period of \textit{Drosophila}'s circadian clock. However, the exact mechanism behind these phenomena is still mostly unknown. Further, reactive oxygen species homeostasis is correlated to the circadian rhythms \cite{Lai2012CCROS,Gyngysi2014,Jimnez2021,Ndiaye2014,Manella2016,deGoede2018,Mezhnina2022}. \par

\begin{figure}
     \begin{subfigure}{0.44\linewidth}
        \hspace*{1.cm}
        \includegraphics[width=1\textwidth]{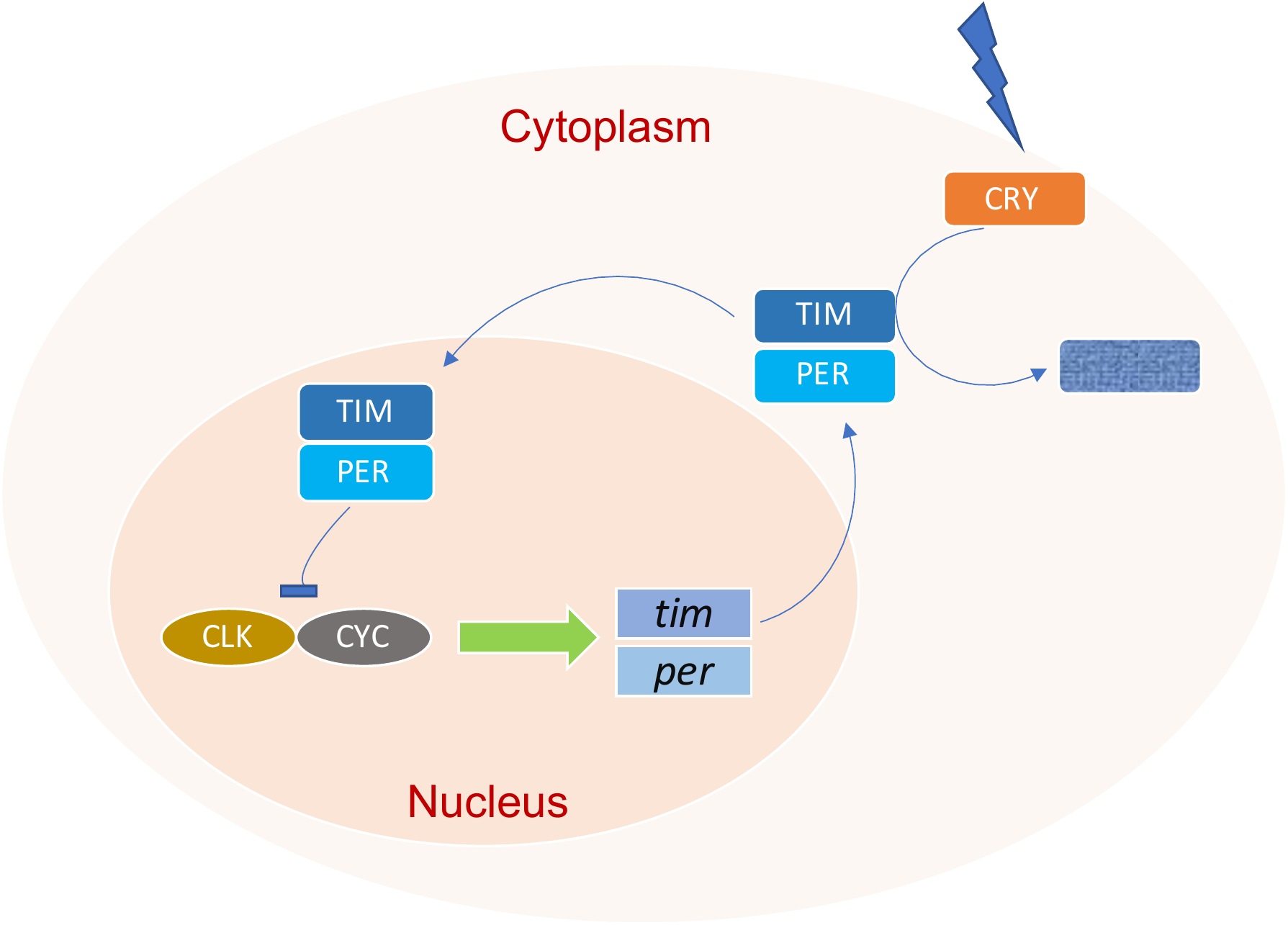}
        \caption{}
        \label{fig:CC-Schem}
    \end{subfigure}
    \hfill
\begin{minipage}{0.4\linewidth}
    \begin{subfigure}{\linewidth}
\hspace*{-1.1cm}
\includegraphics[width=1\textwidth]{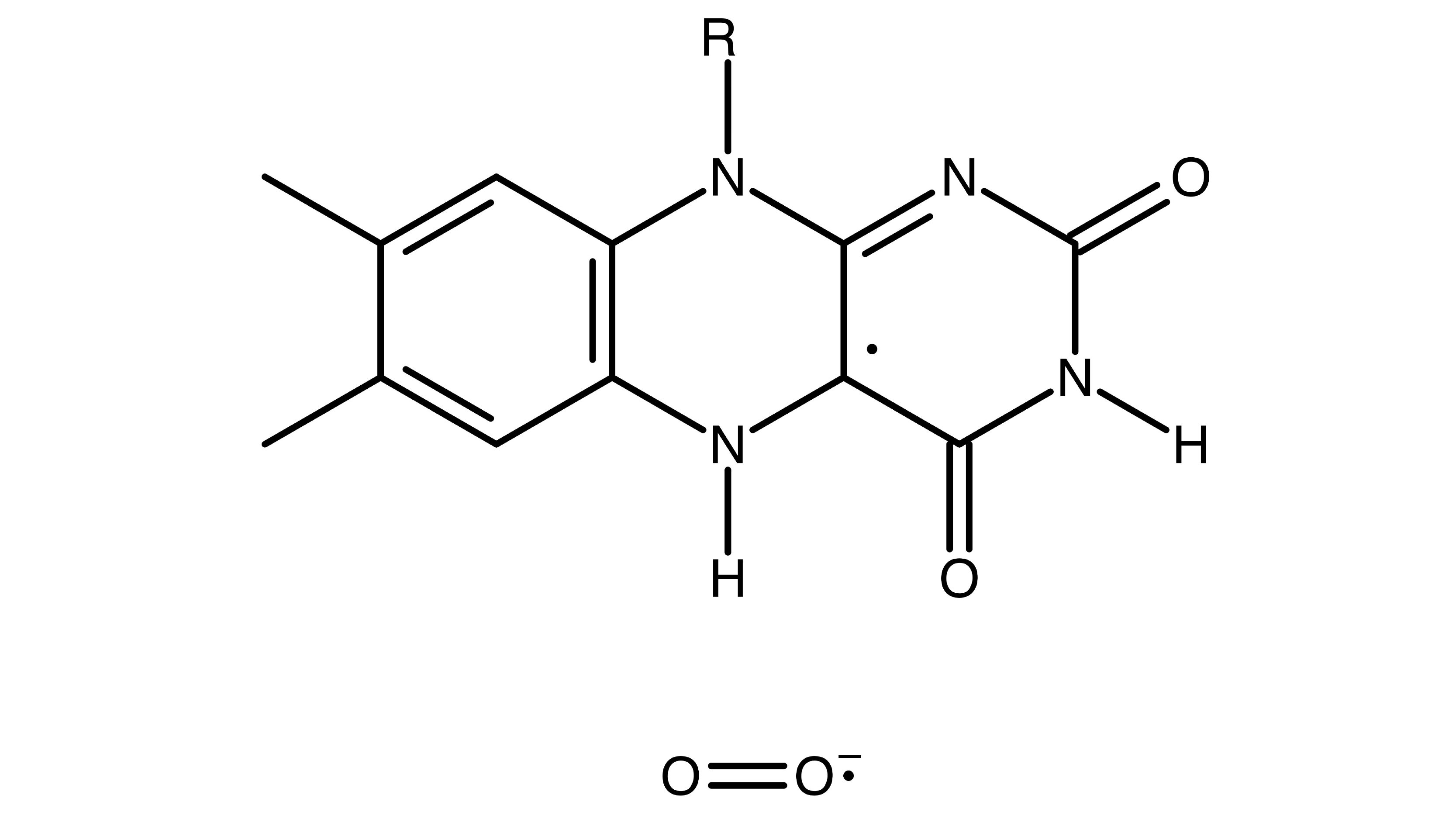}
        \caption{}
        \label{fig:CC-FADH-O2}
    \end{subfigure}
\end{minipage}%
    \hfill
     \begin{subfigure}{0.45\linewidth}
        \hspace*{1.2cm}
        \includegraphics[width=1\textwidth]{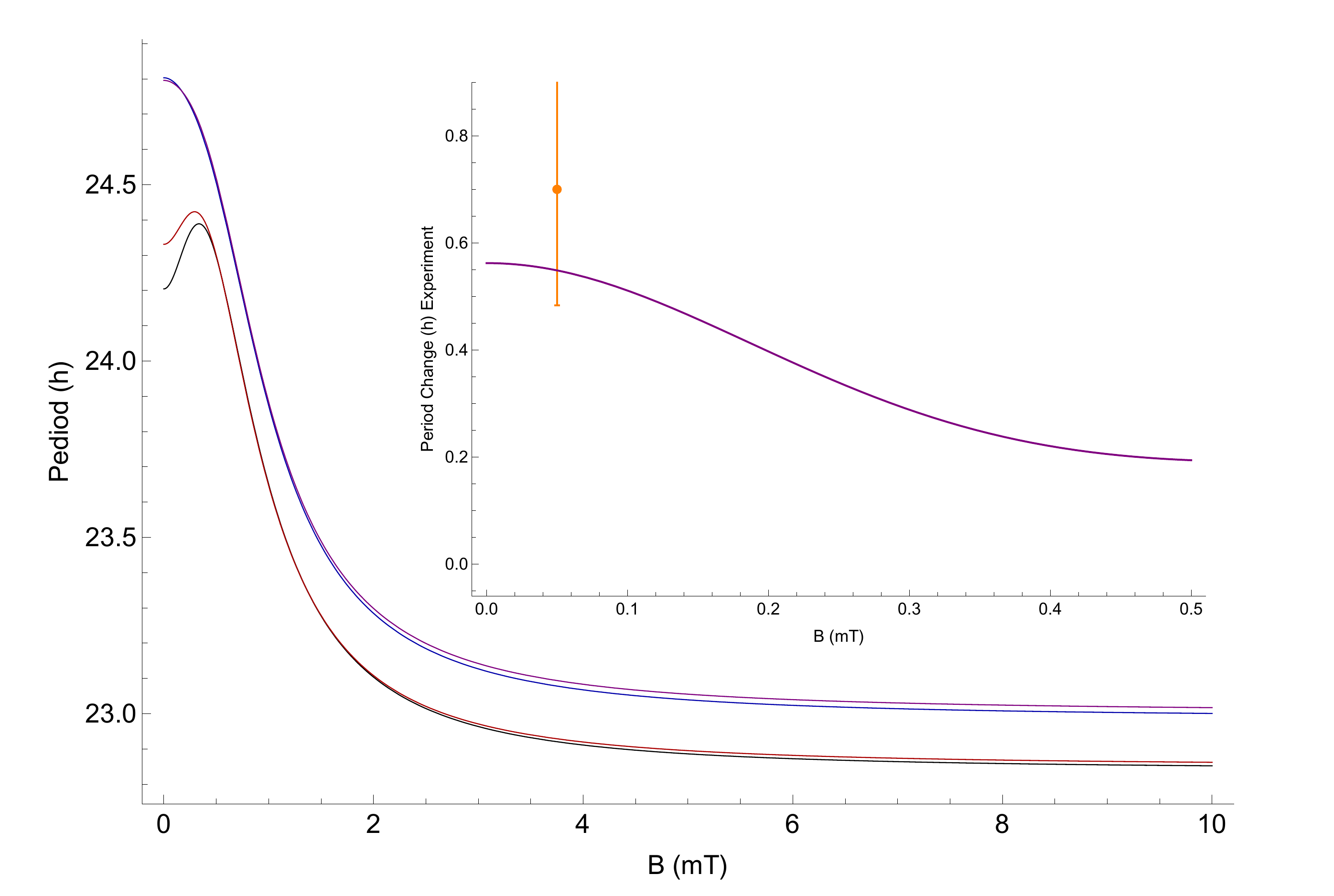}
        \caption{}
        \label{fig:CC-Li}
    \end{subfigure}
    \hfill
\begin{minipage}{0.45\linewidth}
    \begin{subfigure}{\linewidth}
\hspace*{-1.2cm}
\includegraphics[width=1\textwidth]{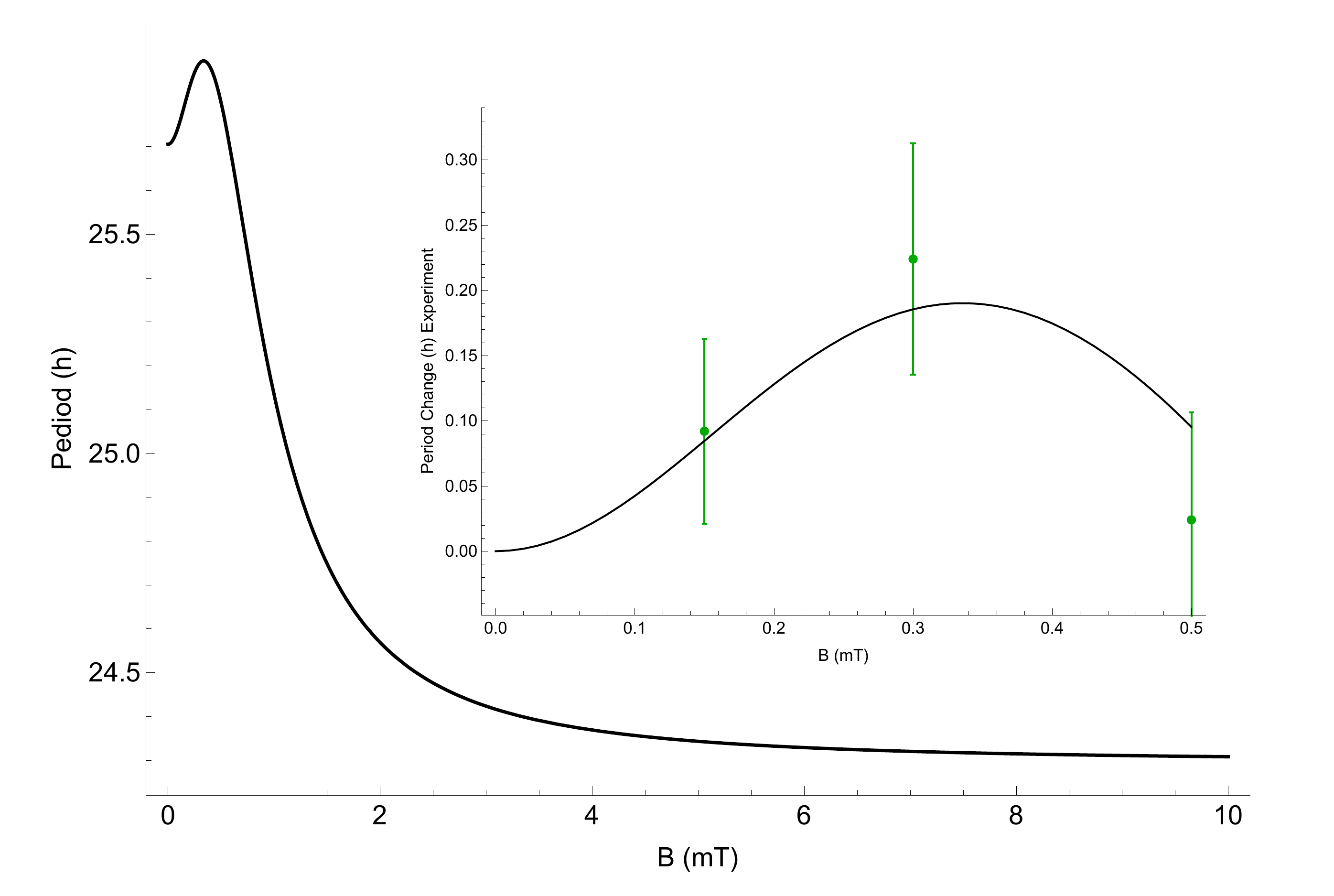}
        \caption{}
        \label{fig:CC-MF}
    \end{subfigure}
\end{minipage}%

\caption{Radical pair explanation for magnetic field and lithium effects on the circadian clock. (a) A simple model of the circadian clock feedback loop in \textit{Drosophila}. CLOCK (CLK) and CYCLE (CYC) proteins promote the $tim$ and $per$ genes. PER and TIM proteins first accumulate in the cytoplasm and then enter into the nucleus to block their gene transcription. Upon light absorption CRY binds to TIM, and this results in the degradation of TIM \cite{Zadeh2022CC}. (b) Flavinsemiquinone (\ch{FADH^{.}}) and superoxide radical pair (\ch{Li^{+}}...\ch{O2^{.-}}). The dependence of the period of \textit{Drosophila}'s circadian clock calculated by the radical pair model on the static magnetic field strength $B$ with (c) and without (d) lithium effects. Higher magnetic field intensities shorten the period of the circadian clock. (c) The effects of \ce{{Li}} [purple], \ce{^6{Li}} [red], \ce{^7{Li}} [blue], and zero \ce{{Li}} [black]. The inset indicates the comparison between the effects of Li on the period of the  clock calculated by the radical pair model [purple line] and the experimental findings [orange dots with error-bars] of Ref. \cite{Dokucu2005}. (d) The comparison between the dependence of the period on applied magnetic field  calculated by the radical pair model [black line in the inset of plot (d)] and the experimental findings [green dots with error-bars] of Ref. \cite{Yoshii2009}. The results from the radical pair model fit the experimental data within the experimental uncertainty.}
\label{fig:CC}
\end{figure}

A recent study suggests that a radical pair model based on [\ch{FADH^{.}}...\ch{O2^{.-}}], similar to Section \ref{sec:RPM-Li}, may explain the magnetic field and lithium effects on \textit{Drosophila}'s circadian clock \cite{Zadeh2022CC}. Following the work of Tyson et al. \cite{Tyson1999}, the authors used a simple mathematical model for \textit{Drosophila}'s circadian clock, as shown in Fig.\ref{fig:CC-Schem} (For more detailed models see Ref \cite{Leloup1999}). Similar to the work of Player et al. \cite{Player2021}, they introduced the effects of applied magnetic fields and hyperfine interactions on the circadian clock process by modifying the corresponding rate representing the role of cryptochrome's light activation and, hence, proteolysis of protein. Based on these models and using Eqs. \ref{eq:SY} and \ref{eq:ham}, they reproduced the experimental findings of magnetic field \cite{Yoshii2009} and lithium effects \cite{Dokucu2005} on \textit{Drosophila}'s circadian clock, as shown in Figs.\ref{fig:CC-Li} and \ref{fig:CC-MF}. The proposed model in that work predicts that lithium influences the clock in an isotope-dependent manner and magnetic fields and hyperfine interactions modulate oxidative stress in the circadian clock.

\subsection{Hypomagnetic field effects on microtubule reorganization} \label{sec:RPM-MT}

Single-cell organisms perform cognitive activities predominantly by cytoskeletal microtubules and are inhibited by anesthetic gases even in the absence of synapses or networks \cite{Craddock2015}. Linganna and colleagues showed that modulation of microtubule stability is a mechanism of action for these anesthetics \cite{Linganna2015}. Bernard reported that anesthetics act directly on cytoplasm, depending on cytoskeletal proteins' dynamics comprising actin filaments and microtubules \cite{Perouansky2012}. Further, Eckenhoff and co-workers found that anesthetics bind to actin and tubulin \cite{Xi2004,Pan2006}. In another study, they show that microtubules play key roles in the action of anesthetics on protein reaction networks involved in neuronal growth, proliferation, division, and communication \cite{Pan2008}. Despite the low binding affinity of anesthetics to tubulin compared to membrane protein, the abundance of tubulin is much more than membrane protein sites. It thus seems plausible that our conscious state of mind is intertwined with microtubules and their dynamics. \par

In recent decades, it has been proposed that quantum physics may explain the mystery of consciousness. In particular the holistic character of quantum entanglement might shed more light on the binding problem \cite{Simon2019}. Penrose and Hameroff proposed that quantum computations in microtubules may be the basis for consciousness \cite{Hameroff2014,Stuart1998,Hagan2002}. It was suggested that electron resonance transfer among tryptophan residues in tubulin in a quantum electronic process could play a role in consciousness \cite{Hameroff2002}. Computational models show that anesthetic molecules might bind in the same regions and hence result in loss of consciousness \cite{Craddock2012}. In a recent work, Zhang and co-workers observed a connection between electronic states and vibrational states in tubulin and microtubules \cite{Zhang2022}. However, quantum electronic coherence beyond ultrafast timescales has been recently challenged experimentally \cite{Cao2020}. In contrast, the coherence of quantum spins can be preserved for much longer timescales \cite{Hu2004}. Similarly, Fisher has proposed that phosphorus nuclear spins could be entangled in networks of Posner molecules which could form the basis of a quantum mechanism for neural processing in the brain \cite{Fisher2015}; however, this sort of spin-based model also demands more supporting evidence \cite{Chen2020}.


A considerable amount of evidence indicates that magnetic fields can influence microtubules \cite{Vassilev1982,Glade2005,Bras1998,Zhang2017elife,Qian2009,Luo2016,Tenuzzo2006,Wu2018}. Wang and colleagues showed that shielding the geomagnetic field caused tubulin assembly disorder \cite{Wang2008}. All these observations point to the magnetosensitivity of microtubules for wide ranges of magnetic field strengths. Further, studies suggest that oxidative stress plays important roles in regulating actin and microtubule dynamics \cite{Wilson2015}. Microtubule contain tryptophan, tyrosine, and phenylalanine residues which are susceptible to oxidation. Further, it is also known that the stability of polymerized microtubules is susceptible to changes in zinc ion concentration in neurons \cite{Craddock2012Zn}.  \par

\begin{figure}
     \begin{subfigure}{0.44\linewidth}
        \hspace*{1.cm}
        \includegraphics[width=1\textwidth]{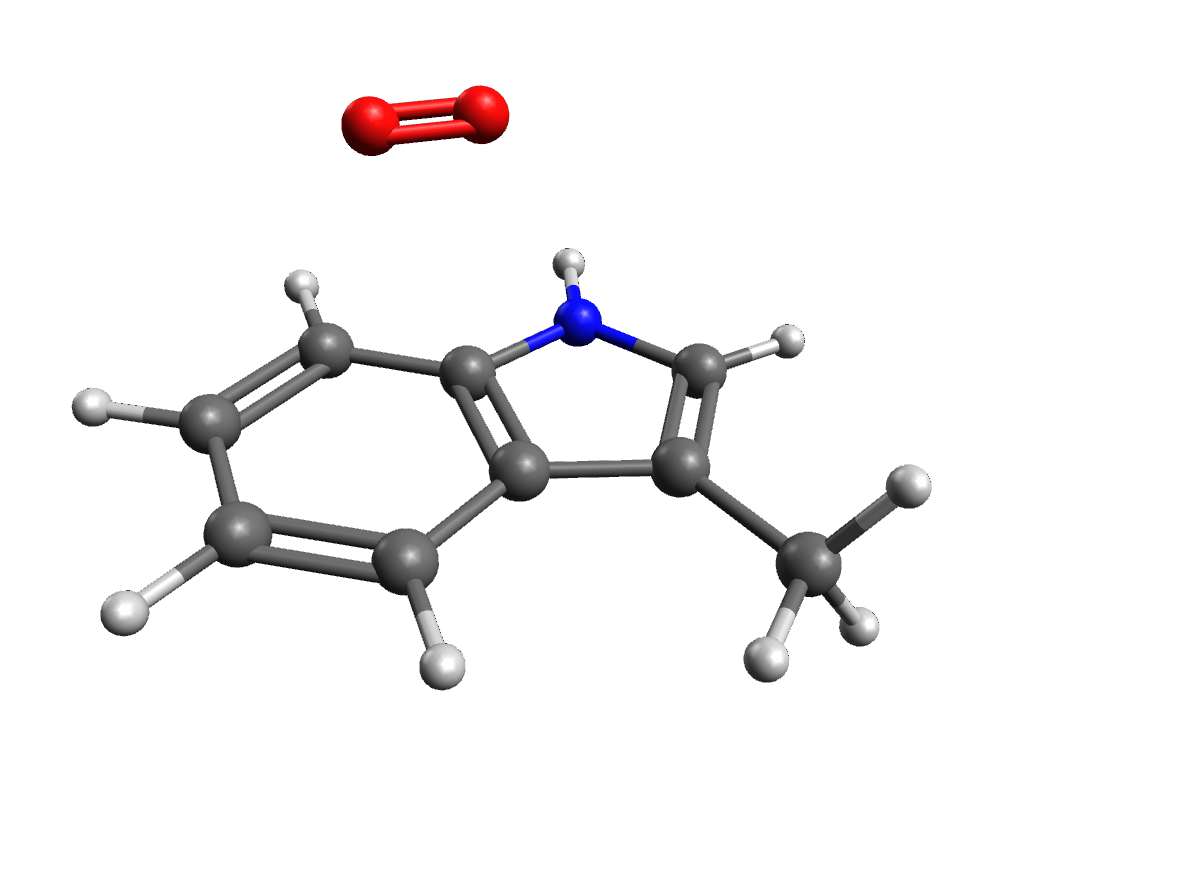}
        \caption{}
         \label{fig:MT-Trp-O2}
    \end{subfigure}
    \hfill
\begin{minipage}{0.4\linewidth}
    \begin{subfigure}{\linewidth}
\hspace*{-1.1cm}
\includegraphics[width=1\textwidth]{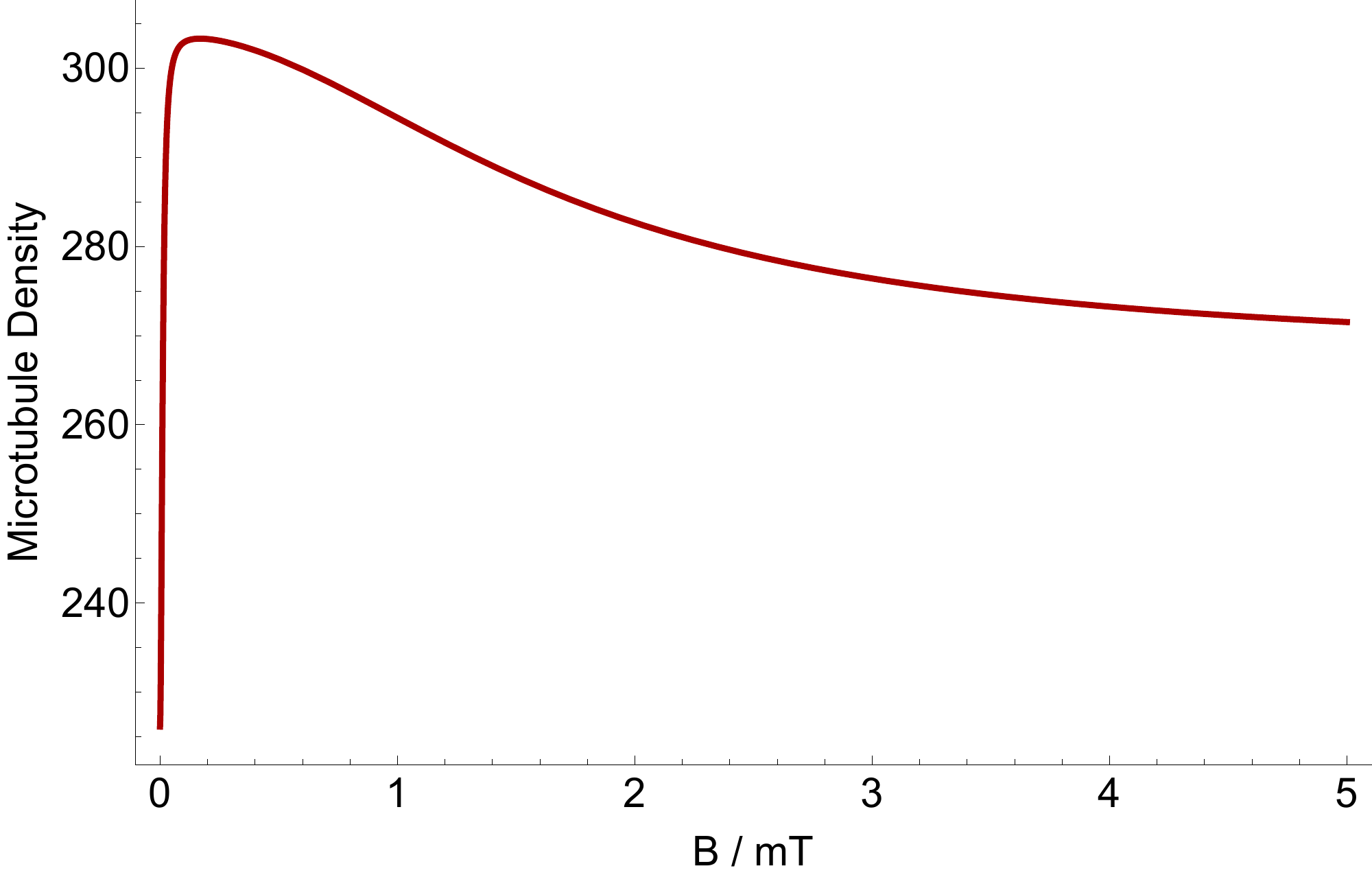}
        \caption{}
       \label{fig:MT-MF}
    \end{subfigure}
\end{minipage}%
    \hfill
     \begin{subfigure}{0.4\linewidth}
        \hspace*{1.1cm}
        \includegraphics[width=1\textwidth]{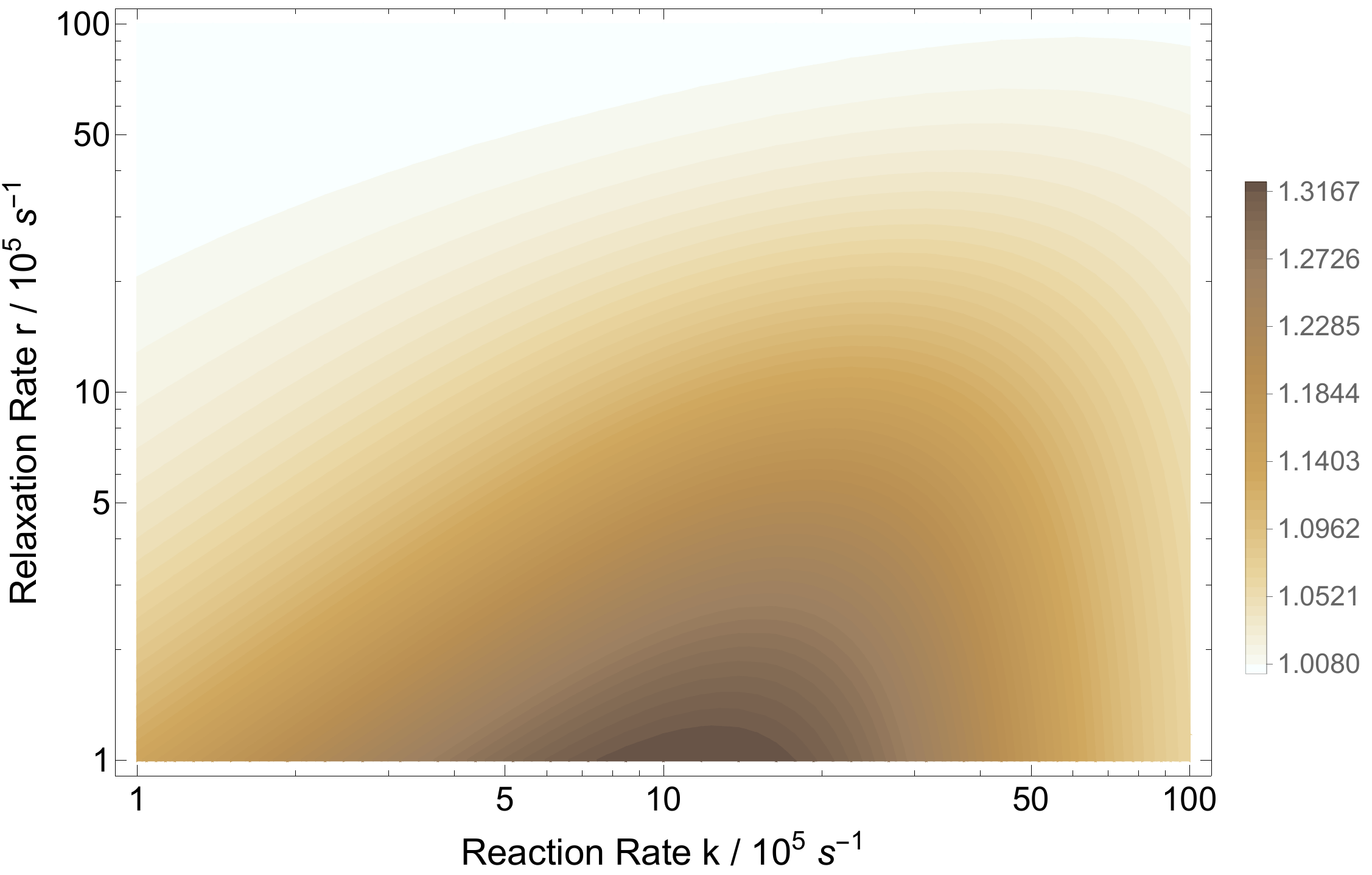}
        \caption{}
        \label{fig:MT-Cntr}
    \end{subfigure}
    \hfill
\begin{minipage}{0.4\linewidth}
    \begin{subfigure}{\linewidth}
\hspace*{-1.1cm}
\includegraphics[width=1\textwidth]{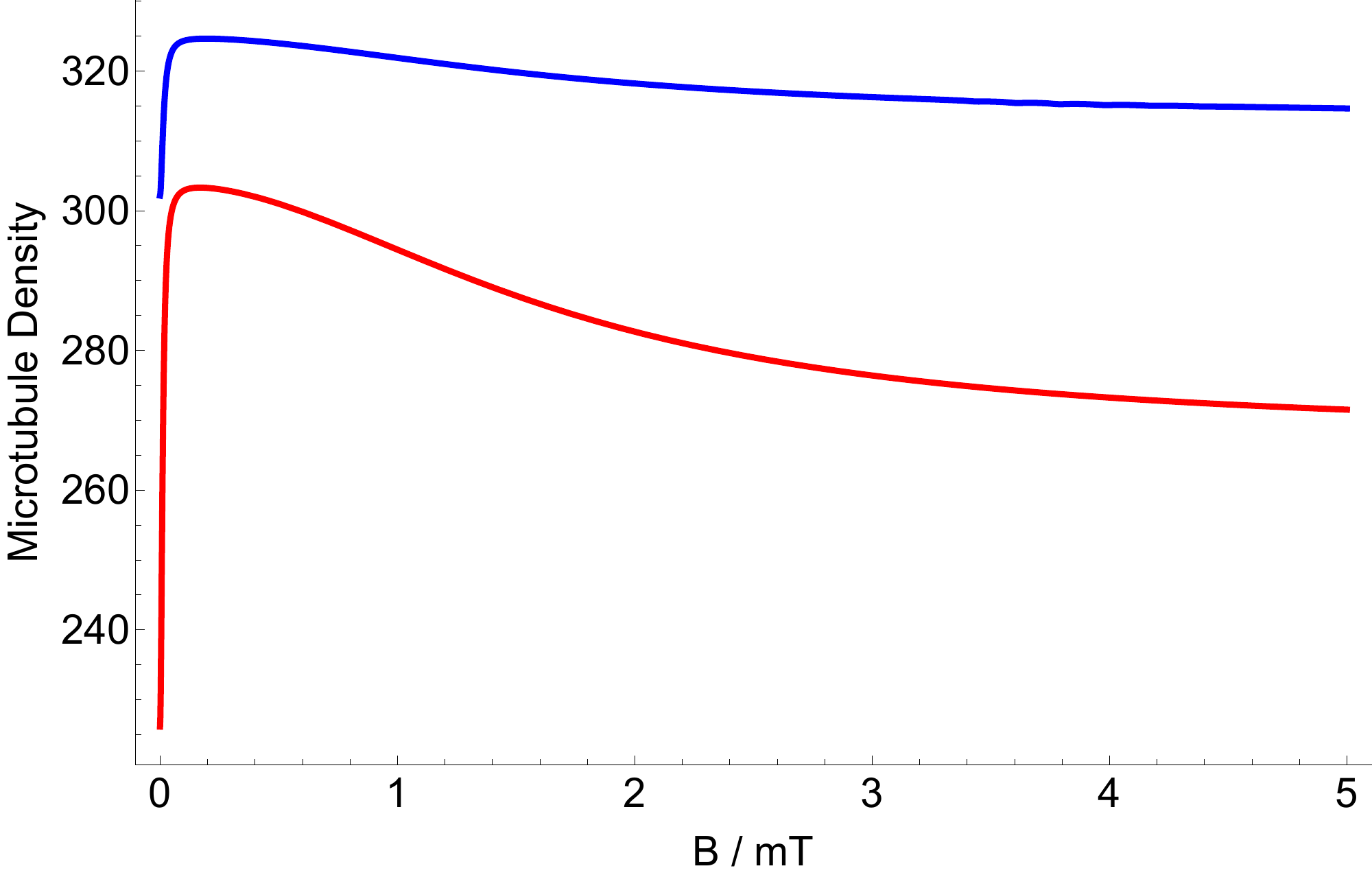}
        \caption{}
         \label{fig:MT-Zn}
    \end{subfigure}
\end{minipage}%

\caption{Radical pair explanation for hypomagnetic field effects on microtubule organization. (a) Schematic presentation of tryptophan ring and superoxide radicals. (b) The dependence of the microtubule density on the applied static magnetic field according to a radical pair model based on [\ch{TrpH^{.+}} ... \ch{O2^{.-}}] complex. The hypomagnetic field causes strong decrease on the microtubule density. The maximum microtubule density occurs around the geomagnetic field. (c) The radical pair model prediction of the microtubule density ratio in geomagnetic field compared to hypomagnetic field. (d) The predicted dependence of the microtubule density on the administration of Zn (with zero nuclear spin) [Red] and \ce{^{67}Zn} (with nuclear spin of $I_{B}=-\frac{5}{2}$) [Blue] as a function of applied magnetic field based on the RP complex of [\ch{TrpH^{.+}} ... \ch{O2^{.-}}] \cite{ZadehHaghighi2022}.}
\label{fig:MT}
\end{figure}

Magnetosensitivty of chemical reactions often involve radical molecules \cite{Rodgers2009} (See also Section \ref{sec:spin-RP}). Using Eqs. \ref{eq:SY} and \ref{eq:ham} and a simple kinetic model \cite{Craddock2012Zn} for dynamics of microtubules, a recent study \cite{ZadehHaghighi2022} suggests that a radical pair model in the form of [\ch{Trp^{.+}}...\ch{O2^{.-}}], similar to Ref. \cite{Smith2021} (See Section \ref{sec:RPM-Xe}), may explain the hypomagnetic field effects on microtubule reorganization reported in Ref. \cite{Wang2008}. They further predict that  the effect of zinc on the microtubule density exhibits isotopic dependence, as shown in Fig. \ref{fig:MT}.

\subsection{Hypomagnetic field effects on neurogenesis} \label{sec:RPM-NG}

In a recent work, Zhang and co-workers showed that shielding the geomagnetic field for a long-term (several weeks) decreased neurogenesis in the hippocampal region in mice \cite{Zhang2021b}. They observed that the neurogenesis impairment was through decreasing adult neuronal stem cell proliferation, altering cell lineages in critical development stages of neurogenesis, impeding dendritic development of newborn neurons in the adult hippocampus, and resulting in impaired cognition. Using transcriptome analysis and endogenous reactive oxygen species \emph{in situ} labeling via hydroethidine, they reported that the hypomagnetic fields reduced levels of reactive oxygen species \cite{Sies2020}. The authors further revealed that such reduction of reactive oxygen species can be compensated by pharmacological inhibition of reactive oxygen species removal via diethyldithiocarbamate, which rescued defective adult hippocampal neurogenesis in hypomagnetic field-exposed mice. \par 
Moreover, it is known that the cellular production of reactive oxygen species is susceptible to the magnetic field exposure \cite{Emre2010,Ghodbane2013,goraca2010effects,Amara2007,Amara2006,Hajnorouzi2011,Kthiri2019,Kamalipooya2017,Politaski2010,Ghodbane2015,Ahn2020,Akdag2010,Amara2010,Cui2012,Chu2012,Zielinski2020,Mert2020,ciejka2011effects,CoballaseUrrutia2018,Manikonda2014}. Reactive oxygen species play vital roles in biology. The mitochondrial electron transport chain and an enzyme family termed NADPH oxidase are two main cellular sources of reactive oxygen species \cite{Sies2020}. The latter is a flavin-containing enzyme. NADPH oxidase enzymes transport electrons from NADPH, through flavin adenine dinucleotide, across the plasma membrane to \ch{O2} to produce \ch{O2^{,-}} \cite{Terzi2020}.\par 

Based on these findings, a recent study \cite{rishabh2021radical} suggests that a radical pair model may explain the modulation of reactive oxygen species production and the attenuation of adult hippocampal neurogenesis in a hypomagnetic field, observed by Zhang and colleagues \cite{Zhang2021b}. The authors proposed that the reduction of the geomagnetic field influences the spin dynamics of the naturally occurring radical pairs in the form of [\ch{FADH^{.}}...\ch{O2^{.-}}], similar to other studies \cite{Player2019,Zadeh2021Li,Zadeh2022CC} (See also Sections \ref{sec:RPM-Li} and \ref{sec:RPM-CC}). They further predict the effects of applied magnetic fields and oxygen isotopic substitution on hippocampal neurogenesis.

 \begin{figure}
        \begin{subfigure}{0.53\linewidth}
            \includegraphics[width=1\textwidth]{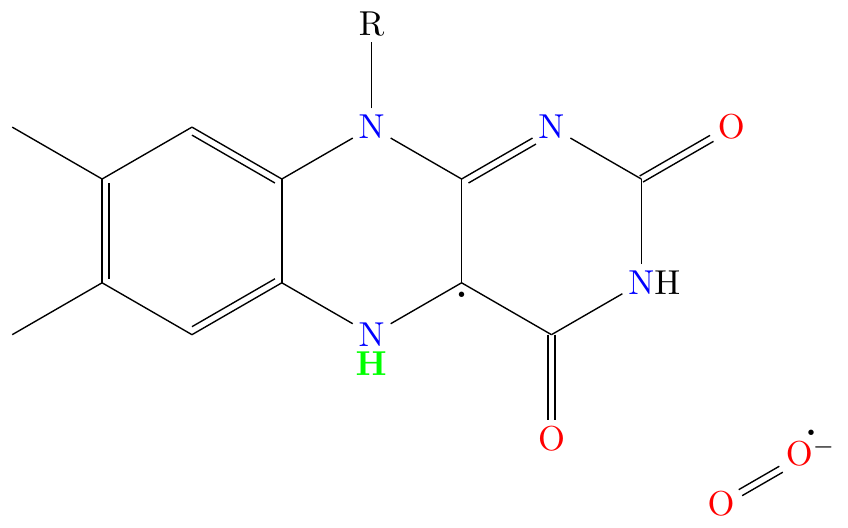}
            \caption{}
            \label{fig:NG-FADH-O2}
     \end{subfigure}
     \hfill
\begin{minipage}{0.53\linewidth}
     \begin{subfigure}{\linewidth}
\includegraphics[width=1\textwidth]{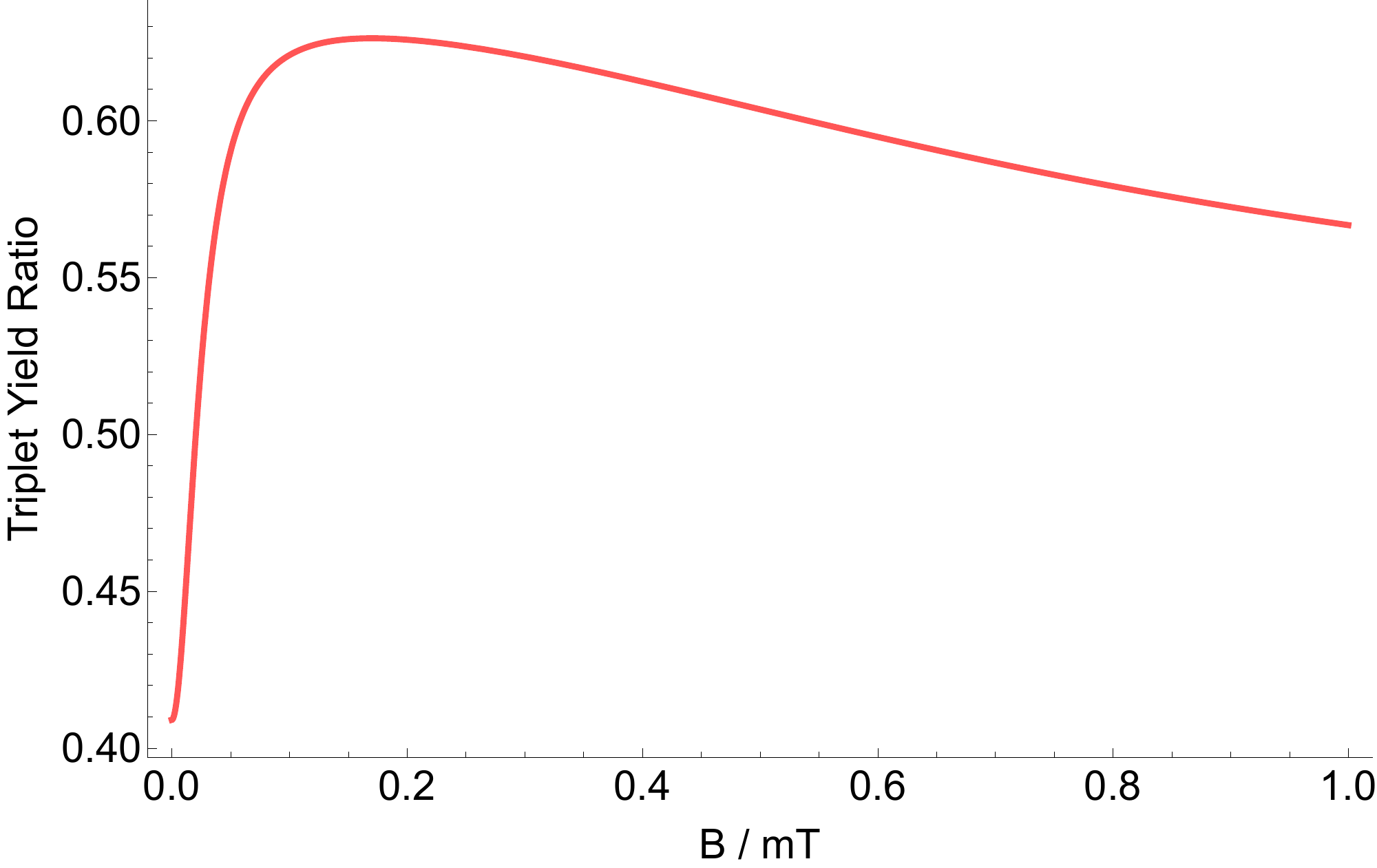}
            \caption{}
            \label{fig:NG-MF}
     \end{subfigure}
\end{minipage}%
     \hfill
        \begin{subfigure}{0.53\linewidth}
             \centering
            \includegraphics[width=1\textwidth]{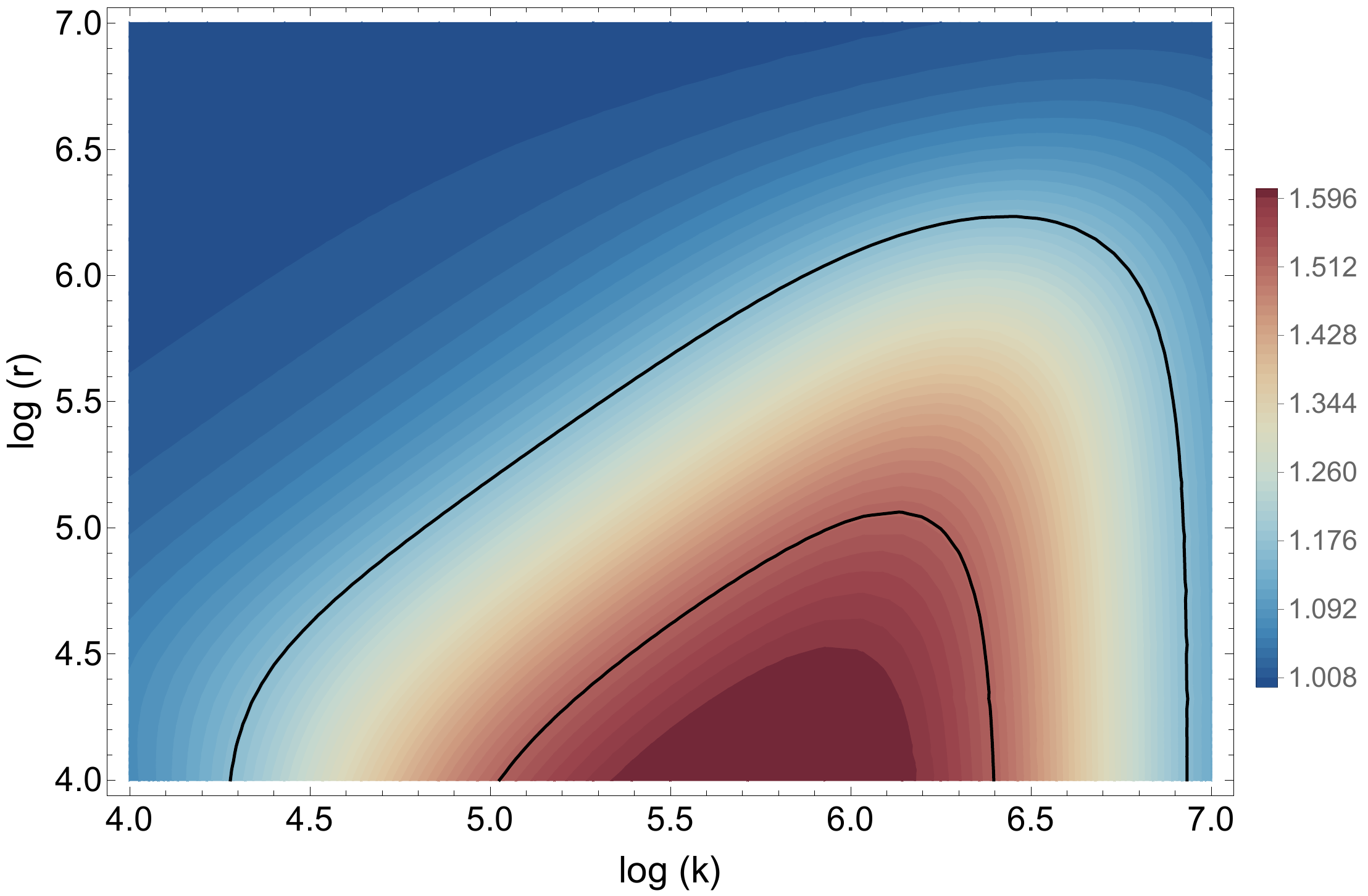}
            \caption{}
            \label{fig:NG-Cntr}
     \end{subfigure}

\caption{Radical pair explanation for hypomagnetic field effects on hippocampal neurogenesis. (a) [\ch{FADH^{.}}...\ch{O2^{.-}}] radical pair. (b) The dependence of the triplet yield of the radical pair model for singlet-born radical pair on external magnetic field \cite{rishabh2021radical}. (c) Triplet yield ratio (geomagnetic field to hypomagnetic field) for singlet-born radical pair in the plan of reaction rate (k) and relaxation rate (r). The region between the solid black lines is in agreement with the experimental range for the ratio of the numbers of BrdU+ cells after an 8 week, observed in Ref. \cite{Zhang2021b}.}
\label{fig:NG}
\end{figure}

\section{Conclusion and outlook} \label{sec:Remarks}

The effects of weak magnetic fields in biology are abundant, including in plants, fungi, animals, and humans. The corresponding energies for such effects are far below thermal energies. So far, there is no explanation for such phenomena. However, quantum biology provides a promising explanation for these effects, namely the radical pair mechanism. Here, we have reviewed numerous studies on the biological effects of weak magnetic fields (static and oscillating), as well as related isotope effects. We then reviewed the radical pair mechanism and proposed that it can provide a unified model for weak magnetic field and isotope effects on biology. We discussed candidate radical pairs that may be formed in biological environments.
We reviewed recent studies that propose that the radical pair mechanism may explain xenon-induced anesthesia, lithium effects on mania, magnetic field and lithium effects on the circadian clock, and hypomagnetic field effects on neurogenesis and microtubule reorganization. These recent studies provide avenues for testing the proposed models. For instance, it is proposed that, in xenon anesthesia, applying magnetic fields $>$ 1mT will increase the anesthetic potency difference between \ch{^{129} Xe} and \ch{^{129} Xe} \cite{Smith2021}. Similarly, it is predicted that for mania treatment by \ch{^{6} Li} and \ch{^{7} Li} \cite{Zadeh2021Li} exposure to hypomagnetic and magnetic fields $>$3 mT will magnify the difference in the potency of these two isotopes. Moreover, it is predicted that exposure of the circadian clock to magnetic fields $>$mT will shorten the period of the clock \cite{Zadeh2022CC}. Another study suggests that exposure to magnetic fields greater than the geomagnetic field will reduce microtubule assembly \cite{ZadehHaghighi2022}. Further, it is also predicted that hippocampal neurogenesis \cite{rishabh2021radical}, the circadian clock \cite{Zadeh2022CC}, and microtubule reorganization \cite{ZadehHaghighi2022} will be isotope-dependent using different isotopes of oxygen, lithium, and zinc, respectively. 

Going beyond these already published proposals, it would be of interest to investigate the roles of radical pairs to help explain magnetic field effects on a large variety of physiological functions, including NMDAR activation \cite{Salunke2013,zgn2019}, DNA/RNA methylation \cite{Baek2019}, dopamine dynamics \cite{Siero2004,janac2009effect}, flavin autofluorescence \cite{Ikeya2021}, epigenetics \cite{Leone2014,Consales2017}, and many others. As discussed earlier in this review, for each of these systems there are naturally occurring radical pairs that can conceivably act as magnetosensitive agents. 
However, in all of the mentioned systems it remains a major open challenge to definitively identify the magnetic sensitive radical pairs as well as the relevant chemical reactions and corresponding kinetic rates. This challenge will require multi-disciplinary collaborations including biologists, chemists, and quantum physicists.  

A considerable amount of evidence indicates that shielding the geomagnetic field has direct biological consequences, which in some cases could be detrimental. This could also be pertinent for the quest of life on other planets without a magnetic field, including Mars \cite{McKay1996,Hyodo2021}. In a similar vein, nowadays almost all species are exposed to magnetic fields at different intensities and frequencies originated by manufactured devices \cite{Khan2021,Burch1999,Maffei2022,Rsli2021,Zastko2021,adair2000static,Touitou2012}. The effects of magnetic fields on physiological functions are inevitable and could be detrimental. Thus this review and perspective is pertinent to the debate on the putative adverse health effects of environmental magnetic fields. Understanding the underlying mechanism should help to clarify many of these issues. \par


From a quantum perspective, it would also be of interest to explore the relevance of quantum entanglement \cite{Wootters1998} in the radical pair models for various magnetic field effects on biological functions \cite{Gauger2011,Pauls2013,Zhang2014}. This could be particularly interesting in the context of neuroscience, where it has been suggested that the brain might use quantum effects such as entanglement for information processing purposes \cite{Hameroff2002,Fisher2015,Kumar2016}.\par



Studying magnetic field and isotope effects in biology is a rich and important interdisciplinary field. The potential essential involvement of quantum effects related to the radical pair mechanism provides an exciting new avenue for further investigation, with the promise of revealing a common underlying mechanism for many of these effects.

\section*{Acknowledgment}

The authors would like to thank Rishabh, D. Salahub, W. Nicola, T. Craddock, A. Jones, M. Ahmad, D. Wallace, A. Lewis, W. Beane, R. Sherrard, M. Lohof, J. Mariani, and D. Oblak for their input. This work was supported by the Natural Sciences and Engineering Research Council of Canada.

\bibliography{sample}

\end{document}